\documentclass[12pt]{report}
\usepackage{mathtools,amssymb,mathrsfs,microtype,color,tikz,geometry,graphicx}
\usepackage[
      colorlinks=true,
      linkcolor=blue,
      urlcolor=blue,
      filecolor=black,
      citecolor=green,
      pdfstartview=FitV,
      linktocpage,
      pdftitle={},
        pdfauthor={x, x, x},
        pdfsubject={},
        pdfkeywords={},
        bookmarksopen=true
      ]{hyperref}

\usepackage{cite}

\makeatletter
\renewcommand{\@dotsep}{10000} 
\makeatother

\makeatletter
\g@addto@macro\bfseries{\boldmath}
\makeatother

\graphicspath{ {figures/} }

\renewcommand{\vec}{\boldsymbol}
\renewcommand{\d}[1]{\widetilde{\mathrm d\boldsymbol{#1}}}
\newcommand{\eqdef}{\overset{\mathrm{def}}=}

\usetikzlibrary{decorations.pathreplacing,angles,quotes}
\usetikzlibrary{arrows,decorations.markings}

\DeclareMathOperator{\diag}{diag}

\newcommand{\remark}[1]
{
\par\bigskip\noindent
\centerline{
\begin{linespread}{1.0} \selectfont
\begin{minipage}{\textwidth}
\small \textbf{Remark:} #1
\end{minipage}
\end{linespread}
}
\par\bigskip
}

\begin{document}

\title{
	{\Huge BMS in higher space-time dimensions and Non-relativistic BMS.}\\\vspace{40pt}
	{\includegraphics[width=.5\textwidth]{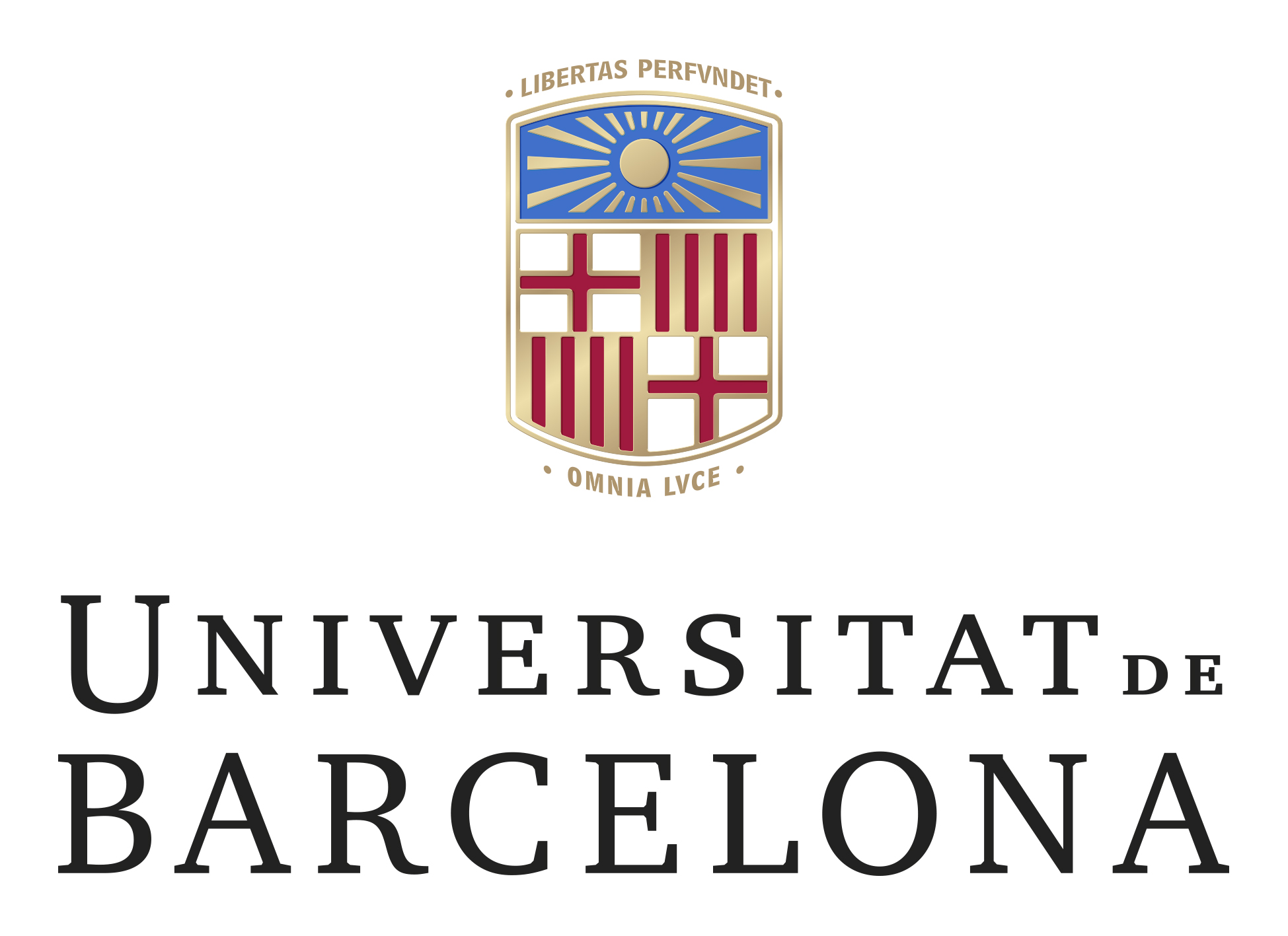}}
}
\author{
Author: \hfill Delmastro, D.G.\hspace{49pt} \\
Supervisors: \hspace{40pt} Batlle, C. and Gomis, J.\\ \vspace{20pt}}
\date{June 2017}
\maketitle

\begin{titlepage}
\vspace*{100pt}
\parbox{.9\textwidth}{
\centering
		{\bfseries \large BMS in higher space-time dimensions and Non-relativistic BMS.}\\[2\baselineskip]
		{\emph{by}}\\
		{Delmastro, D.G.\footnotemark{}}\\
		{\emph{under the supervision of}}\\
		{Batlle, C.\footnotemark{} and Gomis, J.\footnotemark{}}\\[2\baselineskip]
		{Submitted to the department of physics in partial\\ fulfilment of the requirements for the degree of}\\
		{\bfseries Master of science}\\
		{in Astrophysics, Particle Physics and Cosmology}\\
		{at the Universitat de Barcelona.}\\[2\baselineskip]
		\date{June 2017}
}
\footnotetext[1]{email: ddelmade7@alumnes.ub.edu}
\footnotetext[2]{email: carles.batlle@upc.edu}
\footnotetext[3]{email: joaquim.gomis@icc.ub.edu}
\end{titlepage}

{
\tableofcontents
\thispagestyle{empty}
}

\chapter*{Acknowledgments.}
\thispagestyle{empty}

Needless to say, nothing in this document -- or in my career as a whole -- would have been possible without the continuous support of my family. I will forever be grateful to them.

I am also thankful to friends and colleagues, with whom I have shared many priceless conversations and unforgettable experiences, and who not only have helped shape my scientific view of the world but also made the process enjoyable.

In addition, I am also in debt to the many professors I have had the pleasure to interact with and from whom I have learnt invaluable things, both from the Universidad Aut\'onoma de Madrid, where I studied my Bachelor's Degree, and from the Universitat de Barcelona, where I studied my Master's Degree. I would also like to thank the very attentive staff from both universities.

The professors that supervised my final projects in both cycles deserve a special mention. I owe a lot to Antonio Gonz\'alez-Arroyo Espa\~na, from Madrid, and to Carles Batlle Arnau and to Joaqu\'im Gomis Torn\'e, from Barcelona. Last but not least, I am also in debt to the coordinator of the Master's Degree, Bartomeu Fiol N\'u\~nez.

I have refrained from making other names explicit for fear of forgetting someone. I am sure the relevant people will know I am talking about them.

\chapter*{Outline.}\addcontentsline{toc}{chapter}{Outline.}

\setcounter{page}{1}

The Bondi-Metzner-Sachs ($\mathrm{BMS}$) Group is the group of asymptotic isometries of asymptotically flat space-times. This group is given by the semi-direct product of the Lorentz Group with the infinite-dimensional group of the so-called \emph{super-translations}, which constitutes an abelian sub-group.

The isometries of a strictly flat manifold are, as is well-known, Poincar\'e. If one admits a non-trivial geometry, but such that it becomes Minkowski in some of its asymptotic directions, then one may try to find the transformations that leave the asymptotic form of the metric invariant. If the metric is asymptotically flat, the most natural result one would expect is that the asymptotic isometries are Poincar\'e.

\begin{center}
\vspace{-10pt}
\includegraphics[width=.61\textwidth]{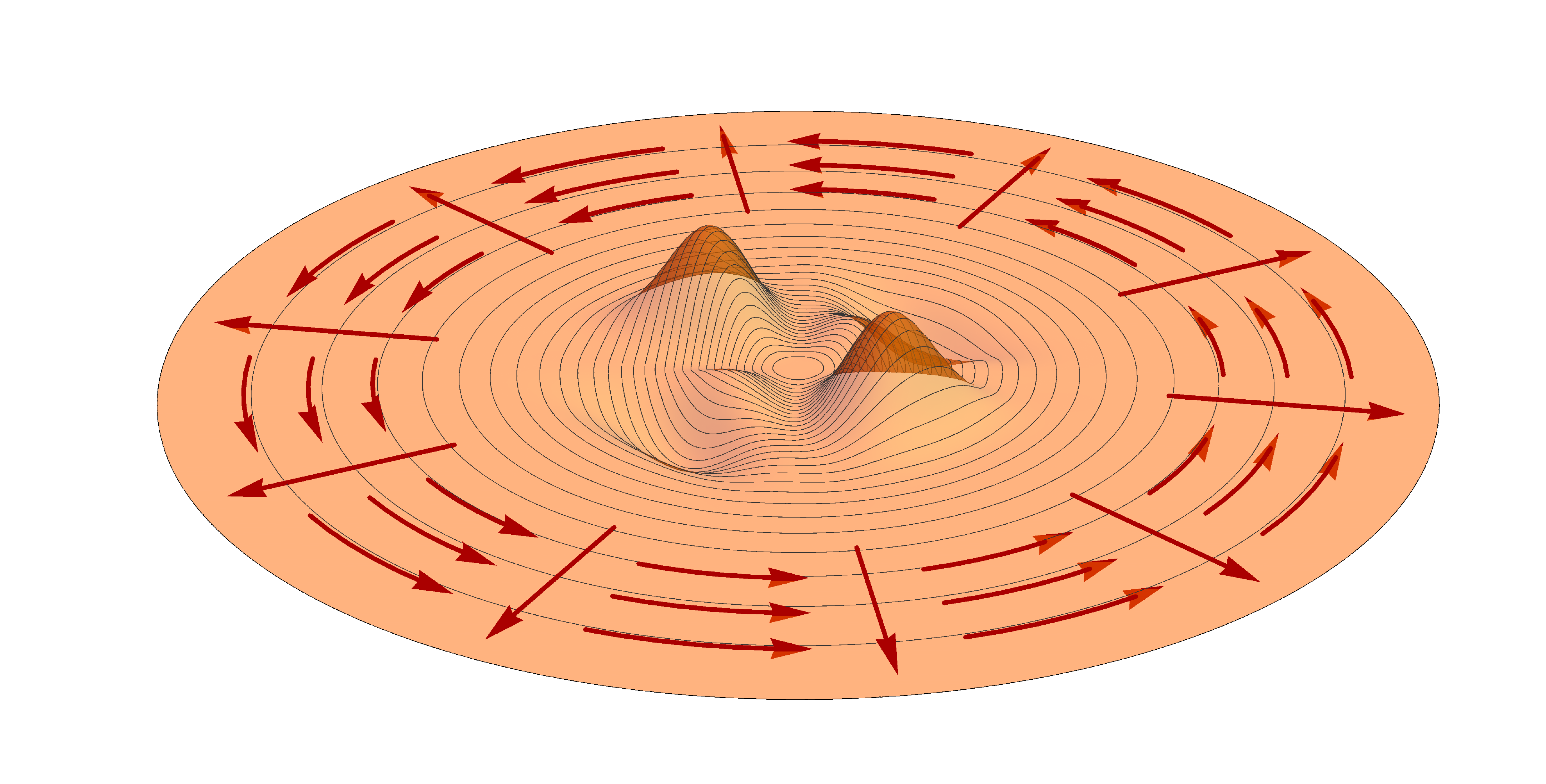}
\vspace*{-15pt}
\end{center}

Indeed, this was the programme that BMS undertook, as a part of their study of gravitational waves, in the early sixties~\cite{1962RSPSA.269...21B,1962RSPSA.270..103S}. In an attempt to rederive the Poincar\'e Group as the group of isometries of a manifold that is asymptotically flat, these authors discovered, much to their surprise, that the actual group of asymptotic isometries does indeed contain Poincar\'e, but it contains much more: it is in fact infinite-dimensional. The new set of symmetries, today called super-translations, can be regarded as a generalisation of the standard translations. We will review the details of the $\mathrm{BMS}$ Group in Chapter~\ref{sec:bms_grav}.

The new physics discovered by these authors soon transcended by far their original intents, to the point that we nowadays consider it to be a just a part of a much larger and richer structure; for a recent review, see the lecture notes~\cite{2017arXiv170305448S}. For example, there is compelling evidence that the $\mathrm{BMS}$ Group is an essential ingredient of quantum gravitational scattering~\cite{bookAshtekar,2014JHEP...07..152S,2015JHEP...05..151H}, black holes and the information paradox~\cite{2015arXiv150901147H,2016PhRvL.116w1301H,2016arXiv161109175H}, flat holography~\cite{2014arXiv1403.3420B,2006RvMaP..18..349D,2004CQGra..21.5655A,2003hep.th....6074B,2003NuPhB.674..553A}, soft theorems~\cite{1965PhRv..140..516W,2014arXiv1411.5745S,2016arXiv161208294C,2016PhRvD..93b6003A}, the memory effect~\cite{1974SvA....18...17Z,2016arXiv161203290H,1991PhRvL..67.1486C}, etc.

The last two of these topics, the soft theorems and the memory effect, have a major role in the study of the $\mathrm{BMS}$ Group. The soft theorems describe the universal behaviour of scattering amplitudes that include massless particles with vanishing momentum. These soft particles can be understood as the Goldstone bosons of spontaneously broken large gauge symmetries; in particular, a broken $\mathrm{BMS}$ symmetry gives rise to soft gravitons. On the other hand, the memory effect is the permanent displacement of two detectors as the consequence of passing-by gravitational waves; in particular, the relative positions and clock times of the detectors before and after the radiation transit differ by a super-translation. As stressed by Strominger and collaborators~\cite{2017arXiv170305448S,2014arXiv1411.5745S}, the memory effect, the soft theorems, and the concept of asymptotic symmetries are the three corners of an \emph{infra-red triangle} -- a triality that is present in any theory with non-trivial infra-red dynamics.

\begin{linespread}{1.0} \selectfont
\begin{center}
\begin{tikzpicture}

\draw [thick,blue,<->,>=stealth] (1,0) -- (3,0);
\draw [thick,blue,<->,>=stealth] (3.5, 0.866) -- (2.5, 2.598);
\draw [thick,blue,<->,>=stealth] (0.5, 0.866) --  (1.5, 2.598);

\node at (0,.2) [align=center] {Soft\\Theorem};
\node at (4,.2) [align=center] {Memory\\Effect};
\node at (2,3.3) [align=center] {Asymptotic\\Symmetry};

\end{tikzpicture}
\end{center}
\end{linespread}
\vspace{-25pt}

This triangular equivalence interweaves these three seemingly unrelated topics and makes them interdependent. This is relevant to our work for the following reason: on the one hand, it has recently been argued that the memory effect vanishes for $d>3$, where $d$ is the number of space dimensions~\cite{2016arXiv161203290H}. On the other hand, the soft theorems exist for any number of space-time dimensions~\cite{2014arXiv1405.3533A}. This puts the third corner, the asymptotic symmetries, in an uncertain position: to what extent should we expect a non-trivial $\mathrm{BMS}$ Group in higher dimensions?

This apparent contradiction has recently been addressed by Strominger et al~\cite{2015arXiv150207644K}. According to these authors, the classical analysis of asymptotic symmetries (e.g.~\cite{2005JMP....46b2503H,2011PhRvD..84d4055T}, etc.) imposed unnecessarily strong conditions for asymptotic flatness, thus eliminating the degrees of freedom corresponding to super-translations. If one slightly relaxes these conditions, the whole family of super-translations emerges, and the paradox concerning the triangle is resolved.

This provides the main motivation for the first part of our thesis (Chapter~\ref{sec:bms_can}): we will argue that one can construct the algebra of the $\mathrm{BMS}$ Group in any number of space-time dimensions, by a different method than the standard approach: we will not consider the gravitational problem, but we will consider instead the \emph{canonical} construction of the $\mathfrak{bms}$ algebra~\cite{1999JMP....40..480L,2017arXiv170301833B,2016PhRvD..93b5030G}. This construction is based on the study of the symmetries of the equations of motion of a free field in flat space-time. As is well-known, the set of point symmetries of such equations contains Poincar\'e transformations; but, as we will show, it is much larger than that, and it contains, in particular, super-translations. The algebra of these symmetries coincides with the one of the gravitational problem, which is one more piece of evidence that suggests that the relevance of the $\mathrm{BMS}$ Group goes beyond the gravitational problem.

The existence of a canonical $\mathrm{BMS}$ Group for any number of space-time dimensions means that either we should expect a non-trivial gravitational $\mathrm{BMS}$ for any $d$, or that the agreement between the canonical and gravitational $\mathrm{BMS}$ in $d=2,3$ is just a fortuitous coincidence. Of course, the first of these two options is not only more attractive, but also much more natural.

The second part of this thesis (Chapter~\ref{sec:nrbms}) concerns the problem of a non-relativistic $\mathrm{BMS}$ Group. One could define this group as the set of isometries of asymptotically flat Newtonian space-times, in the sense of Cartan~\cite{Cartan1923}. Even though some interesting results in this area have been obtained (e.g.,~\cite{2009PhLB..675..133A,2009JHEP...07..037B,2012JHEP...10..092B,2014JPhA...47G5204D}), the overall picture is much less clear than in the relativistic case.

In order to construct the non-relativistic analogue of $\mathfrak{bms}$ there are several approaches one could take. The simplest one is to consider a suitable \.In\"on\"u-Wigner contraction of the corresponding relativistic $\mathfrak{bms}$ algebra, in the same sense that the Galilei algebra can be obtained by contracting the Poincar\'e one. An alternative to obtain a possible non-relativistic $\mathfrak{bms}$ algebra is to mimic the aforementioned canonical construction, but using a scalar field with Galilean space-time symmetries instead of a relativistic field. Finally, a third alternative is, naturally, to follow the same steps that led to the original $\mathfrak{bms}$ algebra, but in a non-relativistic setting, i.e., one could try to study the set of isometries of asymptotically flat Newtonian space-times. In this work, we will only consider the first two possibilities.

It bears mentioning that some partial results concerning the non-relativistic $\mathfrak{bms}_4$ algebra have already been published, cf.~\cite{2017arXiv170503739B}. In this thesis we extend the results therein and contextualise them by comparing them to the relativistic algebra.

\chapter{The $\mathrm{BMS}$ Group.}\label{sec:bms_grav}

In this chapter we will provide a brief but hopefully self-contained introduction to the $\mathrm{BMS}$ Group. A more detailed discussion can be found in~\cite{2017arXiv170305448S,2016arXiv161008526O,2016SchpJ..1133528M,2010JHEP...05..062B}.

The problem that $\mathrm{BMS}$ addressed is the following: given a certain metric $g_{\mu\nu}(x)$ that describes the geometry of an asymptotically flat (lorentzian) manifold,
\begin{equation}
g_{\mu\nu}(x)\xrightarrow{\: x \sim \infty \: }\eta_{\mu\nu},
\end{equation}
we want to find the vector fields $\xi$ such that they leave the asymptotic form of the metric invariant,
\begin{equation}
\mathcal L_\xi g_{\mu\nu}(x)\xrightarrow{\: x \sim \infty \: }0,
\end{equation}
where $\mathcal L$ denotes a Lie derivative (see e.g.~\cite[appendix C]{book:Wald}). Of course, to make this problem well-defined we must properly specify what we mean by the asymptotic limit $x\sim \infty$.

To make the discussion above more precise, let us consider a flat manifold with cartesian coordinates $x^\mu=(x^0,\dots,x^3)$, and let us define the radial coordinate $r$ and the \emph{retarded time} $u_+$ as
\begin{equation}
\begin{aligned}
r&\eqdef\sqrt{x_1^2+x_2^2+x_3^2}\\
u_+&\eqdef x^0-r
\end{aligned}
\end{equation}

The coordinate $u_+$ is the parameter that appears in the solution to Maxwell's equations in the Lorenz gauge~\cite{feynman1964,Jackson1999}, $\partial^2 A^\mu= j^\mu$, viz.
\begin{equation}
\begin{aligned}
A^\mu(x)&=\int \Delta_\mathrm{ret}(x-y) j^\mu(y)\,\mathrm dy\\
&=\int \frac{1}{2\pi}\delta((x_0-y_0)^2-(\boldsymbol x-\boldsymbol y)^2)\Theta(y^0-x^0) j^\mu(y)\,\mathrm dy\\
&=\int \frac{j^\mu(x^0-|\boldsymbol x-\boldsymbol y|,\boldsymbol y)}{4\pi|\boldsymbol x-\boldsymbol y|} \,\mathrm d\boldsymbol y
\end{aligned}
\end{equation}
and as such, it has the following interpretation: if there is a perturbation in the source at a time $u_+$, then a detector placed at a distance $r$ will receive the signal at a time $x^0$. In other words, the world-line of a massless particle is of the form $u_+=\text{const.}$

If we emit massless particles radially away from the origin, at a time $u_+$, then the sphere they reach at null infinity, $r\to\infty$, is called the celestial sphere $\mathcal C\mathcal S^2$. We let $z^A$, with $A=1,2$, parametrise this sphere. Some common choices for these coordinates are: spherical coordinates, $z^A=(\theta,\varphi)$, defined by
\begin{equation}
\begin{aligned}
x^1=r\sin\theta\cos\varphi,\qquad
x^2=r\sin\theta\sin\varphi,\qquad
x^3=r\cos\theta
\end{aligned}
\end{equation}
and the stereographic (complex) coordinates $z^A=(z,\bar z)$, defined by
\begin{equation}
z= \frac{x^1+i x^2}{r+x^3},\qquad \bar z= \frac{x^1-i x^2}{r+x^3}
\end{equation}

We will ultimately be interested in a possible extension of the upcoming discussion to higher dimensional space-times. Therefore, the parametrisation in terms of angles seems to be more convenient than the stereographic one. We will come back to this point later on; for now, we keep the coordinates $z^A$ unspecified.

The coordinates $u_+,r,z^A$ are called \emph{retarded Bondi coordinates}. The set of all celestial spheres,
\begin{equation}
\mathscr{I}^+=\{(u_+,z^A):\ {u_+}\in\mathbb R,\ z^A\in\mathcal C\mathcal S^2\}
\end{equation}
is called \emph{future null infinity}, and it is topologically a cone $\mathbb R\times S^2$.

There exists an analogous construction using \emph{advanced Bondi coordinates}, where one introduces an \emph{advanced time} $u_-\!\eqdef x^0+r$. In the context of electrodynamics, this coordinate appears in the advanced propagator $\Delta_\mathrm{adv}$, and is relevant to, for example, the Wheeler-Feynman absorber theory~\cite{RevModPhys.17.157,RevModPhys.21.425}. 

As in the retarded case, one defines past celestial spheres, which foliate past null infinity $\mathscr{I}^-$. All massless particles originate in $\mathscr{I}^-$ and end up in $\mathscr{I}^+$. On the other hand, massive particles originate in past timelike infinity $i^-$, and end up in future timelike infinity $i^+$. Finally, spatial infinity is denoted by $i^0$ (see Fig.~\ref{fig:penrose_mink}).

\begin{linespread}{1.0} \selectfont
\begin{figure}[!h]
\centering
\begin{tikzpicture}

\begin{scope}[shift={(0,2.5)}]

\draw (-.5,.25) -- (-.13,.45);
\draw (-.5,-.06) -- (-.1,.38);
\draw (-.18,-.13) -- (-.03,.35);
\draw (.18,-.13) -- (.03,.35);
\draw (.5,-.06) -- (.1,.38);
\draw (.5,.25) -- (.13,.45);

\filldraw (0,.5) circle (1.5pt);
\draw[domain=0:360,thick] plot({.7*cos(\x)},{.25*sin(\x)});
\draw[thick] (-.7,0) -- (-.7,-1.5);
\draw[thick] (.7,0) -- (.7,-1.5);
\draw[domain=0:-180,thick] plot({.7*cos(\x)},{-1.5+.25*sin(\x)});
\draw[domain=0:180,dashed] plot({.7*cos(\x)},{-1.5+.25*sin(\x)});
\end{scope}

\filldraw (0,.5) circle (1.5pt);

\draw (-.5,.75) -- (-.13,.55);
\draw (.5,.75) -- (.13,.55);
\draw (-.15,.7) -- (-.06,.63);
\draw (.15,.7) -- (.06,.63);

\draw (-.5,.25) -- (-.13,.45);
\draw (-.5,-.06) -- (-.1,.38);
\draw (-.18,-.13) -- (-.03,.35);
\draw (.18,-.13) -- (.03,.35);
\draw (.5,-.06) -- (.1,.38);
\draw (.5,.25) -- (.13,.45);

\draw[domain=0:360,thick] plot({.7*cos(\x)},{.25*sin(\x)});
\draw[thick] (-.7,0) -- (-.7,-1.5);
\draw[thick] (.7,0) -- (.7,-1.5);
\draw[domain=0:-180,thick] plot({.7*cos(\x)},{-1.5+.25*sin(\x)});
\draw[domain=0:180,dashed] plot({.7*cos(\x)},{-1.5+.25*sin(\x)});

\filldraw (0,-2) circle (1.5pt);

\draw[domain=0:-180,thick] plot({5+2.48*cos(\x)},{.5+.3*sin(\x)});
\draw[domain=0:180,dashed] plot({5+2.48*cos(\x)},{.5+.3*sin(\x)});

\draw[thick] (2.5,.5) -- (5,3) -- (7.5,.5) -- (5,-2) -- (2.5,.5);
\filldraw (5,3) circle (1.5pt);
\filldraw (5,-2) circle (1.5pt);

\begin{scope}[shift={(0,-2.5)}]
\draw (-.5,.75) -- (-.13,.55);
\draw (.5,.75) -- (.13,.55);
\draw (-.15,.7) -- (-.06,.63);
\draw (.15,.7) -- (.06,.63);
\end{scope}

\draw [thick,->,>=stealth] (2.8,3) -- (4.7,3);
\draw [thick,<-,>=stealth] (.3,3) -- (2.2,3);
\node at (2.55,3.05) {$i^+$};

\draw [thick,<-,>=stealth] (.3,.5) -- (.95,.5);
\draw [thick,->,>=stealth] (1.55,.5) -- (2.2,.5);
\node at (1.25,.55) {$i^0$};

\draw [thick,->,>=stealth] (1.6,1.75) -- (1,1.75);
\draw [thick,->,>=stealth] (2.35,1.75) -- (3.3,1.75);
\draw[decoration={brace,mirror},decorate]  (.8,1) -- (.8,2.5);
\draw[decoration={brace},decorate]  (2.5,.7) -- (4.5,2.7);

\node at (1.95,1.75) {$\mathscr I^+$};

\draw [thick,->,>=stealth] (1.6,.75-1.5) -- (1,.75-1.5);
\draw [thick,->,>=stealth] (2.35,.75-1.5) -- (3.3,.75-1.5);
\draw[decoration={brace,mirror},decorate]  (.8,1-2.5) -- (.8,0);
\draw[decoration={brace,mirror},decorate]  (2.5,.3) -- (4.5,-1.7);
\node at (1.95,-0.75) {$\mathscr I^-$};

\draw [thick,->,>=stealth] (2.8,-2) -- (4.7,-2);
\draw [thick,<-,>=stealth] (.3,-2) -- (2.2,-2);
\node at (2.55,-1.95) {$i^-$};

\end{tikzpicture}
\caption{Conformal structure of infinity (Minkowski). Redrawn from~\cite{1963PhRvL..10...66P}.}
\label{fig:penrose_mink}
\end{figure}
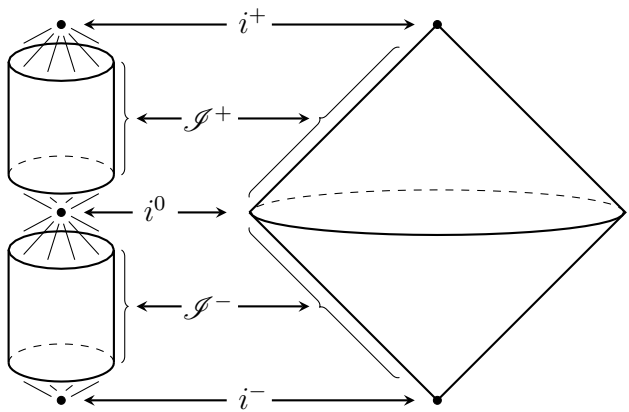
\end{linespread}

In Bondi coordinates, the Minkowski metric reads
\begin{equation}\label{eq:flat_bondi}
\mathrm ds^2=-\mathrm du^2_\pm\mp2\mathrm dr\, \mathrm du_\pm+r^2\gamma_{AB}\mathrm dz^A\mathrm dz^B
\end{equation}
where $\gamma_{AB}$ is the metric on the unit sphere. In spherical coordinates, $\gamma=\diag(1,\sin^2\theta)$, and in stereographic coordinates, $\gamma_{z\bar z}=\gamma_{\bar zz}=\frac{2}{(1+z\bar z)^2}$.

In cartesian coordinates, the Killing vectors of this manifold are given by
\begin{equation}
\xi=a^\mu\partial_\mu+\Omega^{\mu\nu}x_\nu\partial_\mu,
\end{equation}
with $a^\mu,\Omega^{\mu\nu}\in\mathbb R$ a set of constants, with $\Omega^{\mu\nu}=-\Omega^{\nu\mu}$. In Bondi coordinates, these Killing vectors are given by a more complicated and not particularly illuminating expression (see e.g.~\cite{2017arXiv170305448S}). In the particular case of $r\to\infty$, that is, at $\mathscr{I}^\pm$, translations correspond to
\begin{equation}
u\to u+\alpha(z^A)
\end{equation}
where, in spherical coordinates,
\begin{equation}\label{eq:translation_spherical}
\alpha(\theta,\varphi)=a^0+a^1\sin\theta\cos\varphi+a^2\sin\theta\sin\varphi+a^3\cos\theta
\end{equation}

For future reference, one should note that these are just the spherical harmonics $Y_0^0,Y_1^m$, with $m=0,\pm1$. In stereographic coordinates,
\begin{equation}
\alpha(z,\bar z)\sim\text{linear combinations of}\ 1,z,z^2,\bar z,\bar z^2
\end{equation}

On the other hand, (homogeneous) Lorentz transformations correspond to conformal transformations on the celestial sphere, given by
\begin{equation}\label{eq:conformal_lorentz}
\begin{aligned}
z&\to\frac{\omega_{11}z+\omega_{12}}{\omega_{21}z+\omega_{22}}\qquad\text{with}\qquad \omega=\omega(\Omega)\in\mathrm{SL}(2,\mathbb C)\\
u&\to (1+\delta(z,\bar z))u
\end{aligned}
\end{equation}
for a certain function $\delta$ (whose form shall not concern us here).

The infinitesimal generators $M_{\mu\nu}$ of these transformations are given by~\cite{1970JMP....11.3145H}
\begin{equation}
\begin{aligned}
M_{12}&=z\partial_z-\bar z\partial_z\\
M_{13}&=-\frac i2\left((z^2+1)\partial_z+(\bar z^2+1)\partial_{\bar z}\right)\\
M_{23}&=-\frac12\left((z^2-1)\partial_z-(\bar z^2-1)\partial_{\bar z}\right)\\
M_{01}&=+\frac i2\left((z^2-1)\partial_z+(\bar z^2-1)\partial_{\bar z}\right)-iu\frac{z+\bar z}{1+z\bar z}\,\partial_u\\
M_{02}&=+\frac12\left((z^2+1)\partial_z-(\bar z^2+1)\partial_{\bar z}\right)-iu\frac{i(\bar z-z)}{1+z\bar z}\,\partial_u\\
M_{03}&=-i\left(z\partial_z+\bar z\partial_z\right)-iu\frac{1-z\bar z}{1+z\bar z}\,\partial_u
\end{aligned}
\end{equation}

These vector fields satisfy the Lorentz algebra. As a matter of fact, the angular part can be written as linear combination of the generators of the Witt algebra~\cite{1991hep.th....8028G}, $l_n=-z^{n+1}\partial_z,\bar l_n=-\bar z^{n+1}\partial_{\bar z}$, with $n=0,\pm1$.

On the other hand, in spherical coordinates the Lorentz generators are given by
\begin{equation}
\begin{aligned}
M_{12}&=-i\partial_\varphi\\
M_{13}&=i\cos\varphi\,\partial_\theta-i\cot\theta\sin\varphi\,\partial_\varphi\\
M_{23}&=i\sin\varphi\,\partial_\theta+i\cot\theta\cos\varphi\,\partial_\varphi\\
M_{01}&=i\cos\varphi\cos\theta\,\partial_\theta-i\sin\varphi\csc\theta\,\partial_\varphi-iu\cos\varphi\sin\theta\,\partial_u\\
M_{02}&=i\sin\varphi\cos\theta\,\partial_\theta+i\cos\varphi\csc\theta\,\partial_\varphi-iu\sin\varphi\sin\theta\,\partial_u\\
M_{03}&=i\sin\theta\,\partial_\theta+iu\cos\theta\,\partial_u
\end{aligned}
\end{equation}

Upon Lie bracketing, one verifies that these Killing vectors satisfy the Poincar\'e algebra,
\begin{equation}
\begin{aligned}
[P^\mu, P^\nu]&=0\\
[P_\alpha, M^{\mu\nu}]&=2i\delta^{[\mu}{}_\alpha P^{\nu]}\\
[M_{\alpha \beta},  M^{\mu \nu}]&=4i\delta^{[\mu}{}_{[\alpha}   M^{\nu]}{}_{\beta]}
\end{aligned}
\end{equation}


Let us now move on to the non-flat case. With our previous definitions at hand, we we can now properly define what it means for a manifold to be asymptotically flat at $\mathscr I^\pm$: it means that it admits local coordinates $u_\pm,r,z^A$, such that, as $r\to\infty$, the metric takes the form~\eqref{eq:flat_bondi} up to terms subleading in $r$. Apart from the metric becoming flat at null infinity, $g_{\mu\nu}(\mathscr I^\pm)=\eta_{\mu\nu}$, we also assume it to be smooth there. One can prove~\cite{2000CQGra..17.1559T} that this is equivalent to assuming that the components of the metric admit a well-defined $1/r$ expansion around $r\to\infty$ (however, see~\cite{Chrusciel:1993hx}).

\remark{A much more technical and elegant definition of asymptotic flatness can be found in~\cite[chapter 11]{book:Wald} and references therein. This definition is based on the existence of a particular Penrose compactification of the manifold and, as such, it is much more geometrical and coordinate-independent than ours. As discussed in~\cite{2005JMP....46b2503H,2004CQGra..21.5139H}, the Penrose definition is not easily extended to higher (odd) dimensions, due to a potential non-analiticity of the metric at null infinity. On the other hand, the extension to higher dimensions of the Bondi definition in terms of an explicit $1/r$ expansion is in principle much more direct. This latter definition will be enough for our purposes. 
}

For concreteness, we will now focus on the asymptotic future $\mathscr I^+$, so we will drop the `$+$' label on the retarded time. The corresponding analysis at $\mathscr I^-$ is analogous.

Following Bondi, we take $u$ to be a null coordinate, so that $g^{uu}=0$, and we take the angular coordinates $z^A$ to be constant along null rays, so that $g^{uA}=0$. Imposing that the two-spheres spanned by $z^A$ have area $4\pi r^2$, we get $\det g_{AB}=r^4\det \gamma_{AB}$. These conditions fix the diffeomorphism invariance of the metric, and are usually known as the \emph{Bondi gauge conditions}. One should note that some of these conditions are sometimes modified in the literature; for example, Penrose~\cite{1963PhRvL..10...66P} replaces the metric on the sphere, $\gamma_{AB}$ by one conformally equivalent to it, with an arbitrary conformal factor (see~\cite{2000CQGra..17.1559T} for a comparison of the Bondi-Sachs and the Penrose approaches). In any case, we shall stick to the Bondi conditions in what follows.

After fixing the gauge, the metric takes the so-called Bondi-Sachs form~\cite{2016SchpJ..1133528M}:
\begin{equation}
\mathrm ds^2=-\frac{V}{r}\mathrm e^{2\beta}\mathrm du^2-2\,\mathrm e^{2\beta}\mathrm dr\,\mathrm du+r^2h_{AB}\left(dz^A-\frac12 U^A\mathrm du\right)\left(\mathrm dz^B-\frac12 U^B\mathrm du\right)
\end{equation}
where we have defined $h_{AB}\eqdef r^{-2}g_{AB}$, so that the Bondi area gauge condition reads $\det h_{AB}=\det \gamma_{AB}$. The indices on the sphere $A,B,\dots$ are lowered and raised with the metric $h_{AB}$ and its inverse $h^{AB}h_{BC}=\delta^A_C$. We also denote the covariant derivatives with respect to $h_{AB}$ by the letter $D_A$.

Asymptotic flatness, in Bondi coordinates, means
\begin{equation}
\begin{aligned}
\lim_{r\to\infty}\beta&=\lim_{r\to\infty}U^A=0\\
\lim_{r\to\infty}\frac{V}{r}&=1\\
\lim_{r\to\infty}h_{AB}&=\gamma_{AB}
\end{aligned}
\end{equation}

If, as before, we denote the covariant derivative with respect to $h_{AB}$ by $D_A$, and we let $\eth_A$ be the covariant derivative with respect to $\gamma_{AB}$, then the asymptotic condition $h_{AB}\to\gamma_{AB}$ implies that $D_A\to\eth_A$. The precise correspondence is
\begin{equation}
D_Av^B=\eth_A v^B+\frac{1}{2r}\gamma^{BF}\left(\eth_A c_{FE}+\eth_Ec_{FA}-\eth_Fc_{AE}\right)v^E+\mathcal O(r^{-2})
\end{equation}
for an arbitrary vector field $v^B$. Here, $c_{AB}=c_{AB}(z)$ is defined as the subleading part of the metric on the sphere:
\begin{equation}
h_{AB}\eqdef\gamma_{AB}+\frac{c_{AB}}{r}+\mathcal O(r^{-2})
\end{equation}

The rest of the components of the metric must have a well-defined falloff too. There is a certain freedom in the choice of these falloffs, and certain consistency conditions that determine some coefficients in terms of the rest. The (physically motivated) choice for the subleading behaviour of the metric made by Bondi, van der Burg, Metzner and Sachs was
\begin{equation}\label{eq:asymptotic_metric}
\begin{aligned}
V&=r-2M+\mathcal O(r^{-1})\\
\beta&=-\frac{1}{32}\frac{c^{AB}c_{AB}}{r^2}+\mathcal O(r^{-3})\\
U^A&=-\frac{\eth_B c^{AB}}{2r^2}+\frac{2L^A}{r^3}+\mathcal O(r^{-4})
\end{aligned}
\end{equation}
where $M,L$ are arbitrary functions defined on $\mathscr I^+$; $M=M(u,z)$ is called the \emph{mass aspect}, and $L^A=L^A(u,z)$ is called the \emph{angular momentum aspect}.

We are now in position to tackle the problem of characterising the isometries of asymptotically flat space-times; the set of these diffeomorphisms is called the $\mathrm{BMS}$ Group. These are, by definition, the diffeomorphisms that preserve the Bondi gauge conditions as well as the metric falloffs~\eqref{eq:asymptotic_metric}. Preservation of the gauge conditions
\begin{equation}
g_{rr}=g_{rA}=0,\qquad \det h_{AB}=\det \gamma_{AB}
\end{equation}
requires
\begin{equation}\label{eq:gauge_inv}
\mathcal L_\xi g_{rr}=\mathcal L_\xi g_{rA}=\mathcal L_\xi h_{AB}h^{AB}=0
\end{equation}
while preservation of the metric falloffs requires
\begin{equation}
\begin{aligned}\label{eq:falloff_inv}
\mathcal L_\xi g_{ur}=\mathcal O(r^{-2}),\quad \mathcal L_\xi g_{uA}=\mathcal O(1),\quad\mathcal L_\xi g_{AB}=\mathcal O(r),\quad\mathcal L_\xi g_{uu}=\mathcal O(r^{-1})
\end{aligned}
\end{equation}

The general solution to the first set of equations~\eqref{eq:gauge_inv} is
\begin{equation}
\xi(r,u,z)=\alpha(u,z)\partial_u+f^A(u,z)\partial_A+\xi^r\partial_r+\mathcal O(r^{-1})
\end{equation}
where $\alpha,f^A$ are a pair of arbitrary functions, and where the actual form of $\xi^r$ and of the $\mathcal O(r^{-1})$ term shall not concern us here.

With this, the first equation in~\eqref{eq:falloff_inv} implies that
\begin{equation}
\alpha(u,z)=\alpha(z)+\frac12\int_0^{u} \eth_Bf^B(u',z)\,\mathrm du'
\end{equation}
while the second equation implies that $f^B(u,z)=f^B(z)$. The third equation is equivalent to requiring $f^B$ to be  a conformal Killing vector of the sphere,
\begin{equation}
\eth_{(A}f_{B)}-\frac12\gamma_{AB} \eth_Cf^C=0
\end{equation}
while the fourth equation is automatically satisfied.

The infinitesimal generators of the $\mathrm{BMS}$ Group thus read~\cite{2016SchpJ..1133528M}
\begin{equation}
\xi(\mathscr I^+)=\left[\alpha(z)+\frac12u\,\eth_Bf^B(z)\right]\partial_u+f^A(z)\partial_A
\end{equation}
where $\alpha(z)$ is an arbitrary function, and $f^A$ is a conformal Killing vector of the sphere.

The Killing vectors with $\alpha(z)\equiv 0$ correspond to local conformal transformations on the celestial spheres. In complex coordinates, the conformal Killing equation becomes the Cauchy-Riemann equations, which means that $f^z$ (resp.~$f^{\bar z}$) is given by an arbitrary holomorphic (resp.~anti-holomorphic) function on the Riemann sphere $S^2=\mathbb C\cup\{ \infty\}$. One can always write such a function as
\begin{equation}
f^z\partial_z=\sum_{n\in\mathbb Z} f_n l_n,\qquad\text{with}\quad l_n=-z^{n+1}\partial_z
\end{equation}
and a similar relation for $f^{\bar z}$. If $f_n\neq 0$ for $n<-1$, then the Killing vector is singular at $z\to 0$, while if $f_n\neq0$ for $n>1$, then it is singular at $z^{-1}\to 0$. If we restrict ourselves to non-singular functions (so that the global transformation is well-defined) we recover the Lorentz generators~\eqref{eq:conformal_lorentz}, that is, the values $n=0,\pm1$ (for more details, see~\cite{1991hep.th....8028G,2015arXiv150800920O}). It has recently been argued~\cite{2010PhRvL.105k1103B,2011arXiv1102.4632B} that there is no reason to require global well-definedness. If we abandon such condition, the general set of generators is $l_n,\bar l_n$, for $n\in\mathbb N$. The new transformations are called \emph{super-rotations}. We will not be interested in them in this thesis.

\remark{for $d>3$, and due to Liouville's theorem~\cite{book:HertrichJeromin}, the solutions to the conformal Killing equation are just the Lorentz generators, with no super-rotations. A simplified argument for the augmented group of conformal diffeomorphisms on the $d-1=1,2$ spheres is the following~\cite{book:Stoilow}: let $n=d-1$ be the dimension of the celestial spheres $S^n$. Then the conformal Killing equation has $n$ unknowns $f^A$, and $\frac12(n-1)(n+2)$  independent equations. For $n=1$ we get no constraints, so the general solution depends on one arbitrary function. For $n=2$, the two equations for the two unknowns are the Cauchy-Riemann equations, so the general solution depends on two arbitrary functions. For $n> 2$, the system is over-determined, and the general solution has a finite number of real parameters: $\dim\mathfrak{conf}(S^n)<\infty$. Indeed, $\mathfrak{conf}(S^n)$ coincides with the orientation-preserving part of $\mathrm{Spin}(1,d)$, which is the double cover of the Lorentz Group $\text{M\"ob}(n)\cong \mathrm{SO}(1,d)$.}

On the other hand, the vector fields with $f^A(z)\equiv 0$ are called \emph{super-translations}, and they include standard translations for a specific choice of the function $\alpha$. Indeed, if we let
\begin{equation}\label{eq:super_l_01}
\alpha(\theta,\phi)=a^0+a^1\sin\theta\cos\varphi+a^2\sin\theta\sin\varphi+a^3\cos\theta
\end{equation}
we recover equation~\eqref{eq:translation_spherical}, i.e., we find the expected isometries of Minkowski. The surprising fact is that $\alpha(z)$ can be an arbitrary function, which means that the group of isometries is now infinite-dimensional.

\remark{one could argue that Minkowski is itself an asymptotically flat space-time, and therefore its group of asymptotic symmetries includes super-translations as well. In this sense, there would be no enlargement of the group of symmetries when one goes from flat space-time to asymptotically flat space-times: in both cases the group of symmetries is $\mathrm{BMS}$. This is not really the case: the formal definition of an asymptotic symmetry is one such that $\mathcal L_\xi g_{\mu\nu}\approx 0$ ``as good as possible''. In flat space-time, the best approximation to the Killing equation is the Killing equation itself, $\mathcal L_\xi \eta_{\mu\nu}\equiv 0$, whose solutions are exactly Poincar\'e. In non-flat space-times, in general the Killing equation doesn't hold anywhere except possibly at $\mathscr I^\pm$. There, the solutions include an arbitrary function $\alpha$, and there is no general way (or reason) to single out the particular case~\eqref{eq:super_l_01}. If we happen to find translations, they must come hand in hand with super-translations.
}

Following Sachs~\cite{PhysRev.128.2851} (in spherical coordinates) and Barnich and  Troessaert~\cite{2011arXiv1102.4632B} (in stereographic coordinates), we may expand the function $\alpha(z)$ in spherical harmonics $Y(\theta,\varphi)$, or as a Laurent series
\begin{equation}
\begin{aligned}
\alpha(z)&=\sum_{\ell\in \mathbb N}\sum_{m=-\ell}^{+\ell}\alpha^\ell_m\ Y_\ell^m(\theta,\varphi)\\
&=\sum_{n,\bar n\in\mathbb Z}\alpha_{n\bar n}\, z^n\bar z^{\bar n}
\end{aligned}
\end{equation}
which leads to the (countable) infinite-dimensional family of super-translations, generated by
\begin{equation}
\begin{aligned}
P_\ell^m(\mathscr I^+)&\eqdef Y_\ell^m\partial_u\\
P_{n,\bar n}(\mathscr I^+)&\eqdef z^n\bar z^{\bar n}\partial_u
\end{aligned}
\end{equation}

Denoting by $M_{\mu\nu}$ the generators of (homogeneous) Lorentz transformations, one may show that the generators of the $\mathrm{BMS}$ Group satisfy the algebra
\begin{align}
{}[M_{23},P_\ell^m]&=\frac{i}{2} \sqrt{(\ell+m) (\ell-m+1)}\ P_\ell^{m-1}-\frac{i}{2}\sqrt{(\ell-m) (\ell+m+1)}\ P_\ell^{m+1}\notag\\
[M_{31},P_\ell^m]&=\frac{1}{2} \sqrt{(\ell-m) (\ell+m+1)}\ P^{m+1}_\ell+\frac{1}{2} \sqrt{(\ell+m) (\ell-m+1)}\ P^{m-1}_\ell\label{bmsalgebra_grav}\\
[M_{12},P_\ell^m]&=m\, P^m_\ell\notag\\
{}[M_{01},P_\ell^m]&=\frac{\ell +2}{2}\left[ \sqrt{\frac{(\ell-m) (\ell-m-1)}{(2 \ell -1) (2 \ell +1)}}\ P_{\ell-1}^{m+1}+\sqrt{\frac{ (\ell +m)(\ell+m -1)}{(2 \ell -1) (2 \ell +1)}}\ P_{\ell-1}^{m-1}\right]\notag\\
&+\frac{\ell -1}{2}\left[\sqrt{\frac{(\ell +m+1) (\ell +m+2)}{(2 \ell +1) (2 \ell +3)}}\ P_{\ell+1}^{m+1}+\sqrt{\frac{(\ell-m+1)(\ell-m+2) }{(2 \ell +1) (2 \ell +3)}}\ P_{\ell+1}^{m-1}\right]\notag\\
[M_{02},P_\ell^m]&=\frac{i(\ell +2)}{2}\left[  \sqrt{\frac{(\ell-m) (\ell-m-1)}{(2 \ell -1) (2 \ell +1)}}\ P_{\ell-1}^{m+1}-\sqrt{\frac{ (\ell +m)(\ell+m -1)}{(2 \ell -1) (2 \ell +1)}}\ P_{\ell-1}^{m-1}\right]\notag\\
&+\frac{i(\ell -1)}{2}   \left[\sqrt{\frac{(\ell +m+1) (\ell +m+2)}{(2 \ell +1) (2 \ell +3)}}\, P_{\ell+1}^{m+1}-\sqrt{\frac{(\ell-m+1)(\ell-m+2) }{(2 \ell +1) (2 \ell +3)}}\, P_{\ell+1}^{m-1}\right]\notag\\
[M_{03},P_\ell^m]&=i (\ell +2) \sqrt{\frac{(\ell -m) (\ell +m)}{(2 \ell -1) (2 \ell +1)}}\ P_{\ell-1}^m-i (\ell -1)\sqrt{\frac{(\ell -m+1) (\ell +m+1)}{(2 \ell +1) (2 \ell +3)}}\ P_{\ell+1}^m\notag
\end{align}
or, in stereographic coordinates,
\begin{equation}
\begin{split}
[l_n,l_{n'}]&=(n-n')l_{n+n'}\\
[\bar l_{\bar n},\bar l_{\bar n'}]&=(\bar n-\bar n')\bar l_{\bar n+\bar n'}
\end{split}\qquad
\begin{split}
[l_p,P_{n,\bar n}]&=\left(\tfrac{1}{2}(p+1)-n\right)P_{n+p,\bar n}\\
[\bar l_p,P_{n,\bar n}]&=\left(\tfrac{1}{2}(p+1)-\bar n\right)P_{n,\bar n+p}
\end{split}
\end{equation}

This is the $\mathfrak{bms}_4$ algebra. In stereographic coordinates, the result is much more compact and transparent, but it has the disadvantage that its generalisation to higher dimensions is not straightforward. We shall henceforth focus on the parametrisation in terms of spherical coordinates.

One of the most prominent features of the $\mathfrak{bms}_4$ algebra is that it contains, as one would expect, a Poincar\'e sub-algebra. Indeed, if we restrict ourselves to the super-translations with $\ell=0,1$, then the algebra above is easily seen to contain a closed sub-algebra, because the coefficients of the terms with $\ell\to\ell+1$ are proportional to $(\ell-1)$, which vanishes at $\ell=1$. Under the identification
\begin{equation}
\begin{split}
P^1&=\frac{P^{1}_1+P^{\bar 1}_1}{i\sqrt{2}}\\
P^2&=\frac{P_1^1-P_1^{\bar 1}}{\sqrt{2}}\\
P^3&=P^{0}_1
\end{split}\qquad\overset{\text{inverse}}{\Longleftrightarrow}\qquad
\begin{split}
P_1^{\bar 1}&=\frac{-P^2+iP^1}{\sqrt{2}}\\
P_1^1&=\frac{+P^2+iP^1}{\sqrt{2}}\\
P^{0}_1&=P^3
\end{split}
\end{equation}
one may check that the $\ell=0,1$ sub-set of the $\mathfrak{bms}_4$ algebra coincides with the Poincar\'e algebra. This in turns means that there is a well-defined notion of energy and momentum for asymptotically flat manifolds; this is the Bondi momentum $p_B$~\cite{1998PhRvD..58h4001C}.

\remark{the Killing equation that we used to determine the super-translations is homogeneous, which means that the normalisation of $P_\ell^m$ is in principle arbitrary. Here we have followed the standard convention $P_\ell^m=Y_\ell^m\partial_u$, which we shall call the \emph{gravitational normalisation}. Had we included some $\ell$-dependent coefficient in front of the spherical harmonics, the actual value of the structure constants of the algebra~\eqref{bmsalgebra_grav} would have been different. In the next chapter we will rederive the $\mathfrak{bms}$ algebra in a different context, and there we will find that the most natural normalisation for the super-translations is different; we will call it the \emph{canonical normalisation}. These two normalisations will be related through an $\ell$-dependent rescaling.}

To extend the discussion above to higher-dimensional space-times, we would have to make a careful analysis of the falloff conditions, and repeat the calculation of the Killing vector fields. From a  more pragmatic and perhaps na\"ive point of view, we could simply anticipate that the solution takes the same form as in the four dimensional case,
\begin{equation}
\xi(\mathscr I^+)\overset?=\left[\alpha(z)+\frac{u}{d-1}\eth_Bf^B(z)\right]\partial_u+f^A(z)\partial_A
\end{equation}
where $\alpha(z)$ is an arbitrary function, and $f$ is a conformal Killing vector on $S^{d-1}$,
\begin{equation}
\eth_{(A}f_{B)}-\frac{1}{d-1}\gamma_{AB}\, \eth_Cf^C=0
\end{equation}

It has recently been argued~\cite{2016arXiv161203290H} that in $d>3$ the most natural falloff conditions for the metric are actually too strong to admit a non-trivial $\mathrm{BMS}$ Group. In such a case, the only admissible solutions to the $\mathrm{BMS}$ problem would be Poincar\'e, that is, the Killing vectors $\xi$ would correspond to Lorentz transformations and standard translations. The case is not closed yet though:~\cite{2015arXiv150207644K} counter-argues that these conditions are unnecessarily strong, and that the $\mathrm{BMS}$ Group exists and is non-trivial in any number of space-time dimensions.

In any case -- be it physically realistic or not -- we can dismiss these arguments and impose whatever falloff conditions we may need in order for $\xi(\mathscr I^+)=\alpha(z)$ to be a valid asymptotic isometry. 
In this case, the calculation of the $\mathfrak{bms}_{d+1}$ algebra becomes straightforward: we just need to expand the super-translations in (higher dimensional) spherical harmonics, $Y_\ell$, where $\ell=(\ell_1,\ell_2,\dots,\ell_{d-1})$ is a multi-index:
\begin{equation}\label{eq:killing_d}
P_\ell(\mathscr I^+)\eqdef Y_\ell\,\partial_u
\end{equation}
and calculate the different Lie brackets among the $P_\ell,M_{\mu\nu}$. We will come back to this problem, from a different perspective, in Chapter~\ref{sec:bms_can}. There we will see that the $\mathfrak{bms}$ algebra is a natural construction in any number of space-time dimensions, which suggests that, irrespective of its applicability to the gravitational problem, the $\mathrm{BMS}$ Group exists and is non-trivial for any $d$.

To close this chapter, it is worthwhile mentioning that the group that we have been studying is $\mathrm{BMS}^+$, corresponding to the symmetries that act on future null infinity $\mathscr I^+$. The analogous group acting on past null infinity $\mathscr I^-$ is denoted by $\mathrm{BMS}^-$. Both groups admit extensions in the sense of super-rotations (and even an extension to include conformal transformations~\cite{2017arXiv170108110H}). Moreover, it has recently been conjectured~\cite{2014JHEP...07..152S} that the so-called diagonal subgroup of $\mathrm{BMS}^+\!\times\mathrm{BMS}^-$ is the group of symmetries of (both classical and quantum) gravitational scattering, giving rise to an infinite number of conservation laws. This subgroup is obtained by a certain identification of future-acting to past-acting super-translations, $\alpha^+(z)=\alpha^-(z)$. We refer the reader to~\cite{2014JHEP...07..152S,2017arXiv170305448S} for more details.

\chapter{Canonical realisation of $\mathrm{BMS}$.}\label{sec:bms_can}

In this chapter we will  study a subset of the symmetries of the equations of motion of a relativistic free field $\phi$ of arbitrary spin in flat space-time. This object can be regarded as a classical field, or as a quantum operator; for definiteness, here we will only consider the first case, though our main conclusions will hold in the second case as well.

Let the equations of motion of $\phi:\mathbb R^{d+1}\to\mathbb C^s$  be
\begin{equation}
\mathscr D\phi=0
\end{equation}
for $\mathscr D$ a certain Poincar\'e invariant linear differential operator. By Poincar\'e invariance we mean that $\mathscr D$ commutes with translations and Lorentz transformations:
\begin{equation}
[\mathscr D,-i\partial_\mu]=[\mathscr D,ix_{[\nu}\partial_{\mu]}+\text{spin}]=0
\end{equation}
where ``spin'' corresponds to a finite-dimensional matrix that generates the representation of the Lorentz Group under which $\phi$ transforms. In simple terms, Poincar\'e invariance means that $\mathscr D$ does not depend on $x$, and that its spin indices are contracted in a Lorentz-invariant way. Typical examples include the Klein-Gordon equation, the Dirac equation or the Maxwell equations.

What we want to do is to find the on-shell symmetries of $\mathscr D$, that is, the differential operators $Q$ such that
\begin{equation}
[\mathscr D,Q]\propto \mathscr D
\end{equation}

It is clear that if $-i\partial_\mu$ is a symmetry of $\mathscr D$, then so is $f(-i\partial_\mu)$ for any function $f$. Standard translations are recovered from $f$ by taking it to be the identity. We will see that, for a certain family of functions $f$, the associated differential operators $f(-i\partial_\mu)$ will correspond to super-translations, and they will therefore furnish a representation of the $\mathfrak{bms}$ algebra. The point of this construction is that it is qualitatively independent of the number of space-time dimensions, so it provides a natural extension of the $\mathfrak{bms}_4$ algebra to any $d$.

\remark{instead of considering $\phi$ to be a free field, we could take it to be an interacting field such that the interactions vanish at large times $t\to\pm\infty$. In fact, this is the usual setting considered in interacting quantum field theories, where fields are asymptotically free (e.g., the LSZ formalism~\cite{1955NCimS...1..205L} or the more technical Haag-Ruelle scattering theory~\cite{1958PhRv..112..669H,QFTRuelle}, the Gell-Mann and Low theorem~\cite{1951PhRv...84..350G}, etc.). This would lead to two groups of symmetries, at $t\to\pm\infty$, and one could perhaps draw an analogy with $\mathrm{BMS}^\pm$, acting on $\mathscr I^\pm$.}

The equation $\mathscr D\phi=0$, being free, can be solved most easily in momentum space. We will frame our discussion directly in terms of $a,a^*$, that is, the Fourier modes of $\phi$. From now on, no explicit reference to the underlying field $\phi$ shall be made (we will come back to it in Chapter~\ref{sec:nrbms}).

Let us therefore introduces a pair of canonical variables $a_{\vec k,\sigma},a_{\vec k,\sigma}^*$, with algebra
\begin{equation}
\begin{aligned}
\{a_{\vec k,\sigma},a_{\vec k',\sigma'}\}&=0\\
\{a_{\vec k,\sigma},a^*_{\vec k',\sigma'}\}&=\delta_{\sigma\sigma'}\delta_{\vec k\vec k'}
\end{aligned}
\end{equation}
where $\{\cdot,\cdot\}$ denotes a Dirac bracket.

Here, $\vec k\in\mathbb R^d$ is a continuous index -- the momentum -- and $\sigma\in\mathbb N$ is a discrete index -- the spin. On the other hand, $\delta_{\sigma\sigma'}$ is the Kronecker delta, while $\delta_{\vec k\vec k'}$ is the Dirac delta on the mass hyperboloid,
\begin{equation}
\delta_{\vec k\vec k'}\eqdef (2\pi)^d2\sqrt{\vec k^2+m^2}\ \delta(\vec k-\vec k')
\end{equation}


Given this pair of canonical variables, one may realise the Poincar\'e algebra through the standard construction,
\begin{equation}
\begin{aligned}
 P_\mu&\eqdef\sum_\sigma\int a^*_{\vec k,\sigma}k_\mu a_{\vec k,\sigma}\,\d k\\
 M_{0i}&\eqdef t P_i+\sum_\sigma\int a^*_{\vec k,\sigma} \mathcal M_{0i} a_{\vec k,\sigma}\,\d k+\text{spin}\\
 M_{ij}&\eqdef\sum_\sigma\int a^*_{\vec k,\sigma} \mathcal M_{ij} a_{\vec k,\sigma}\,\d k+\text{spin}
\end{aligned}
\end{equation}
where $\d k$ is the invariant measure on the mass hyperboloid,
\begin{equation}
\d k\eqdef\frac{1}{(2\pi)^d}\frac{\mathrm d\vec k}{2\sqrt{\vec k^2+m^2}}
\end{equation}
and $\mathcal M_{\mu\nu}$ are the differential operators
\begin{equation}
\begin{aligned}
\mathcal M_{0i}&\eqdef ik_0\partial_i\\
\mathcal M_{ij}&\eqdef 2ik_{[i}\partial_{j]}
\end{aligned}
\end{equation}

The differential operators $\mathcal M_{\mu\nu}$ satisfy the Lorentz algebra,
\begin{equation}
[ \mathcal M_{\alpha \beta},\mathcal   M^{\mu \nu}]=4i\delta^{[\mu}{}_{[\alpha} \mathcal  M^{\nu]}{}_{\beta]}
\end{equation}
while the functions $P_\mu,M_{\mu\nu}$ satisfy the Poincar\'e algebra,
\begin{equation}
\begin{aligned}
\{ P^\mu, P^\nu\}&=0\\
\{ P_\alpha, M^{\mu\nu}\}&=2i\delta^{[\mu}{}_\alpha P^{\nu]}\\
\{ M_{\alpha \beta},  M^{\mu \nu}\}&=4i\delta^{[\mu}{}_{[\alpha}   M^{\nu]}{}_{\beta]}
\end{aligned}
\end{equation}

\remark{if we regard $\mathcal M_{\mu \nu}$ as a vector field instead of a differential operator, then it is straightforward to check that it satisfies the Killing equation, that is, $\mathcal M_{\mu \nu}$ generates the isometries of the mass hyperboloid which, as a manifold (Euclidean $\mathrm{AdS}_d$), is maximally symmetric. This corresponds to the well-known isomorphism $\mathrm{Iso}(\mathrm{EAdS}_d)=\mathrm{SO}(1,d)$.}

If we think of the functions $a,a^*$ as the Fourier modes of a free field, then the objects $ P_\mu, M_{\mu\nu}$ are actually the Noether charges corresponding to space-time symmetries. In particular, $ P_\mu$ are the generators of space-time translations, and they are conserved. In fact, for free theories one may actually construct an infinite number of conserved charges,
\begin{equation}
Q_f=\sum_\sigma\int a^*_{\vec k,\sigma}f(\vec k) a_{\vec k,\sigma}\,\d k
\end{equation}
where $f$ is an arbitrary function defined on the mass hyperboloid. For each $f$ we will have a charge which, albeit perhaps non-local, is conserved $\dot Q_f=0$.

The standard Poincar\'e algebra is obtained by taking $f(\vec k)= k^\mu$, in which case the conserved charge $Q_f$ coincides with the generators of translations, $ P^\mu$. The generalisation of the Poincar\'e algebra to the $\mathfrak{bms}$ algebra is obtained by extending the functions $k^\mu$ into the larger set $\chi_\ell^\lambda(\vec k)$,
\begin{equation}
\begin{aligned}
k^\mu&\to m\chi^\lambda_\ell(\vec k)\\
P^\mu&\to P^\lambda_\ell\eqdef m\sum_\sigma\int a^*_{\vec k,\sigma}\chi_\ell^\lambda(\vec k) a_{\vec k,\sigma}\,\d k
\end{aligned}
\end{equation}
for a certain set of (dimensionless) functions $\chi^\lambda_\ell$ defined on the mass hyperboloid. Here, $\ell=(\ell_1,\ell_2,\dots,\ell_{d-1})\in\mathbb Z^{d-1}$ is a certain multi-index whose precise role will be clarified below; while $\lambda\in\mathbb R$ is a real parameter. The set of indices $(\lambda,\ell)$ generalise the vector index $\mu$.

We will call the symmetries generated by the conserved charges $P_\ell^\lambda$ \emph{super-translations}. As long as the set of functions $\{\chi^\lambda_\ell\}$ contains $k^\mu$ for certain values of $(\lambda,\ell)$, the corresponding conserved charges will agree with the generators of standard translations. Furthermore, if the functions $\{M_{\mu\nu},P_\ell^\lambda\}$ satisfy a closed algebra, then we will have succeeded in finding a generalisation of the Poincar\'e algebra. The new algebra, the $\mathfrak{bms}$ algebra, will be the sought-after infinite-dimensional generalisation of the Poincar\'e algebra. In what follows, we will describe how the functions $\chi^\lambda_\ell$ are constructed.

The functions $ P^\lambda_\ell$ clearly commute among themselves, while the functions $ M_{\mu\nu}$ satisfy the Lorentz algebra. The $\mathfrak{bms}$ algebra is obtained by extending the relation $\{ P_\alpha, M^{\mu\nu}\}=2i\delta^{[\mu}{}_\alpha P^{\nu]}$ into
\begin{equation}
\{ P^\lambda_\ell, M_{\mu\nu}\}=\sum_{\alpha\in\mathbb Z^{d-1}}c^\alpha_{\mu\nu}\, P^\lambda_{\ell+\alpha}
\end{equation}
for a certain set of structure constants $c^\alpha_{\mu\nu}=c^\alpha_{\mu\nu}(\lambda,\ell)$. Equivalently, we can also write the relation above as
\begin{equation}\label{eq:BMS_algebra}
\mathcal M_{\mu\nu}\chi^\lambda_\ell=\sum_{\alpha\in\mathbb Z^{d-1}}c^\alpha_{\mu\nu}\,\chi^\lambda_{\ell+\alpha}
\end{equation}

From this equation we see that what we want  is essentially to construct an infinite-dimensional representation of the Lorentz algebra. Therefore, we will define the functions $\chi^\lambda_\ell$ as the eigenvectors of the Casimir operator of the Lorentz algebra,
\begin{equation}\label{eq:defDelta}
\Delta\eqdef\frac12 \mathcal M_{\mu\nu} \mathcal M^{\mu\nu}
\end{equation}
such that
\begin{equation}
\Delta\chi^\lambda_\ell=\lambda\chi^\lambda_\ell
\end{equation}

For one thing, we note that the functions $k^\mu$ themselves satisfy $\Delta k^\mu=d\,k^\mu$, which nicely fits in our picture: we are trying to generalise the functions $k^\mu$ into a larger set, so it is reassuring to see that they indeed satisfy the equation we are using to define $\chi^\lambda_\ell$.

In order to better understand the equation above, it is useful to point out that $\Delta$ is actually the Laplace-Beltrami operator on the mass hyperboloid $k^2+m^2=0$. In other words, given the flat metric $\mathrm ds^2=-\mathrm dk_0^2+\mathrm d\boldsymbol k^2$ and the embedding $k^0\to +\sqrt{\vec k^2+m^2}$, the induced metric is $\mathrm ds^2=g_{ij}\mathrm dk^i\mathrm dk^j$, where
\begin{equation}
g_{ij}=\delta_{ij}-\frac{k^ik^j}{\vec k^2+m^2}\qquad\overset{\text{inverse}}{\Longleftrightarrow}\qquad g^{ij}=\delta_{ij}+\frac{k^i k^j}{m^2}
\end{equation}
with determinant $g=\det(g_{ij})=\frac{m^2}{\vec k^2+m^2}$.

With this, the induced Laplacian reads
\begin{equation}
\frac{1}{m^2}\Delta=\frac{1}{\sqrt g}\partial_i(\sqrt gg^{ij}\partial_j)=g^{ij}\partial_i\partial_j+\frac{d}{m^2} k^i\partial_i
\end{equation}
which agrees with~\eqref{eq:defDelta}. For completeness, we also mention that the Ricci scalar of this manifold is $R=-d(d-1)/m^2$ and the Christoffel symbols are $\Gamma^k_{ij}=-g_{ij}k^k/m^2$.

With this, we will define our functions $\chi^\lambda_\ell$ to be the eigenvectors of $\Delta$, that is, the solutions of
\begin{equation}\label{eq:def_omega_l}
\Delta\chi^\lambda_\ell=\lambda\chi^\lambda_\ell
\end{equation}

The operator $\Delta$ is formally self-adjoint, and has an empty point spectrum~\cite{1966JMP.....7.2026L}. In what follows we will be interested in the continuous spectrum of $\Delta$ and, in particular, in the singular value $\lambda=d$, as we shall discuss later on. For this particular value of $\lambda$, the equation above becomes invariant under conformal transformations $g_{ij}\to \Omega^2 g_{ij}$, $\chi\to \Omega^{(2-d)/2}\chi$.

The equation~\eqref{eq:def_omega_l} does not determine $\chi^\lambda_\ell$ uniquely, because the eigenvalues $\lambda$ are degenerate; this is where the labels $\ell$ come into play. In order to fix the functions $\chi^\lambda_\ell$, we will simultaneously diagonalise the set of commuting operators $\{\Delta, \mathcal M_{12},\Delta_{S^n}\}$, where $\Delta_{S^n}$ is the Laplacian of the unit $n$-sphere,
\begin{equation}
\Delta_{S^n}=\sum_{n>i>j} \mathcal M_{ij} \mathcal M^{ij}
\end{equation}
whose eigenvalues are of the form $-\ell_n(\ell_n+n-1)$, with $\ell_n\in\mathbb N$ a non-negative integer (see the appendix~\ref{sec:spherical_harmonics}).

With this, our functions $\chi^\lambda_\ell$ are defined through
\begin{equation}\label{eq:def_omega}
\left\{\begin{aligned}
\Delta\chi^\lambda_\ell&=\lambda\chi^\lambda_\ell\\
 \mathcal M_{12}\chi^\lambda_\ell&=\ell_1\chi^\lambda_\ell\\
\Delta_{S^n}\chi^\lambda_\ell&=-\ell_n(\ell_n+n-1)\chi^\lambda_\ell\qquad n=2,3,\dots,d-1
\end{aligned}\right.
\end{equation}
with $|\ell_1|\le\ell_2\le \ell_3\le\cdots\le\ell_{d-1}$. The last of these labels, $\ell_{d-1}$, plays the role of ``total angular momentum'', so we will sometimes denote it by $L\eqdef \ell_{d-1}$.

The equations~\eqref{eq:def_omega}, together with univaluedness and finiteness, are enough to uniquely determine the functions $\chi^\lambda_\ell$. We will solve them next. These equations are separable in spherical coordinates $(\rho,\theta)$:
\begin{equation}
\begin{aligned}
k^1&=\rho\sin\theta_{d-1}\sin\theta_{d-2}\cdots\sin\theta_2\sin\theta_1\\
k^2&=\rho\sin\theta_{d-1}\sin\theta_{d-2}\cdots\sin\theta_2\cos\theta_1\\
k^3&=\rho\sin\theta_{d-1}\sin\theta_{d-2}\cdots\cos\theta_2\\
&\cdots\\
k^{d-1}&=\rho\sin\theta_{d-1}\cos\theta_{d-2}\\
k^d&=\rho\cos\theta_{d-1}\\
\end{aligned}
\end{equation}

In these coordinates, the Laplacian reads
\begin{equation}
\frac{1}{m^2}\Delta=\left(1+\frac{\rho^2}{m^2}\right)\partial_\rho^2+\left(\frac{d-1}{\rho}+\frac{d\,\rho}{m^2}\right)\partial_\rho+\frac{1}{\rho^2}\Delta_{S^{d-1}}
\end{equation}
where $\Delta_{S^{d-1}}$ is the Laplacian on the $(d-1)$-sphere. The equations
\begin{equation}\label{eq:mode_angular_eig}
\begin{aligned}
\mathcal M_{12}\chi^\lambda_\ell&=\ell_1\chi^\lambda_\ell\\
\Delta_{S^n}\chi^\lambda_\ell&=-\ell_n(\ell_n+n-1)\chi^\lambda_\ell\qquad n=2,3,\dots,d-1
\end{aligned}
\end{equation}
determine the angular part of $\chi(\rho,\theta)$ to be the spherical harmonics $Y_\ell$,
\begin{equation}
\chi^\lambda_\ell(\rho,\theta)=f(\rho)Y_\ell(\theta),
\end{equation}
while the radial equation, as given by
\begin{equation}
\left(\Delta-\lambda\right)\chi(\rho,\theta)=0
\end{equation}
reads
\begin{equation}
\left(1+z^2\right)f''+\left(\frac{d-1}{z}+d\,z\right)f'-\left(\frac{L(L+d-2)}{z^2}+\lambda\right)f=0
\end{equation}
where $z=\rho/m$, and a prime denotes differentiation with respect to $z$.

Through  the change of variables $f(z)=z^L F(-z^2)$, this equation adopts the standard hypergeometric form~\cite[eq.~15.5.1]{AbramowitzStegun}
\begin{equation}
z\left(z\frac{\mathrm d}{\mathrm dz}+\alpha\right)\left(z\frac{\mathrm d}{\mathrm dz}+\beta\right)F=z\frac{\mathrm d}{\mathrm dz}\left(z\frac{\mathrm d}{\mathrm dz}+\gamma-1\right)F
\end{equation}
where
\begin{equation}
\begin{aligned}
\alpha&=\frac{1}{4} \left(+\sqrt{(d+1)^2+4 (\lambda -d)}+d-1+2 L\right)\\
\beta&=\frac{1}{4} \left(-\sqrt{(d+1)^2+4 (\lambda -d)}+d-1+2 L\right)\\
\gamma&=\frac{d}{2}+L
\end{aligned}
\end{equation}

The solution to the radial equation thus reads
\begin{equation}\label{eq:radial_f_12}
\begin{aligned}
f^\lambda_L(\rho)&=c_1 \left(\frac{\rho}{m}\right)^L  {}_2F_1\left[\frac{d+L+\Lambda-1}{2},\frac{L-\Lambda}{2};\frac{d}{2}+L;-\frac{\rho ^2}{m^2}\right]\\
&+c_2\left(\frac{m}{\rho}\right)^{d+L-2} {}_2F_1\left[\frac{\Lambda-L+1}{2},\frac{2-d-\Lambda-L}{2};2-\frac{d}{2}-L;-\frac{\rho ^2}{m^2}\right]
\end{aligned}
\end{equation}
where $_2F_1$ is the Gaussian hypergeometric function, and
\begin{equation}
\Lambda\eqdef \frac{1}{2}\left (\sqrt{(d+1)^2+4 (\lambda-d)}-d+1\right)\qquad\overset{\text{inverse}}{\Longleftrightarrow}\qquad \lambda=\Lambda (\Lambda+d-1)
\end{equation}

As $L\ge 0$, the solution corresponding to $c_2$ is singular at $\rho\to 0$, so we will only keep the one corresponding to $c_1$ (which we set to $c_1\equiv 1$, which we dub the canonical normalisation). For $\lambda<0$ the function $f_L^\lambda(\rho)$ decays as $\rho\to\infty$, while for $\lambda\ge0$ it does not. From now on we will focus on the second case.

Using ${}_2F_1(\alpha,\beta;\gamma;z)=1+\mathcal O(z)$, together with~\cite[eq.~15.3.7]{AbramowitzStegun}
\begin{equation}
\begin{aligned}
{}_2F_1(\alpha ,\beta ;\gamma ;-z)&=z^{-\alpha }\frac{\Gamma (\gamma )  \Gamma (\beta -\alpha )  }{\Gamma (\beta ) \Gamma (\gamma -\alpha )}\ {}_2F_1\left(\alpha ,\alpha -\gamma +1;\alpha -\beta +1;-\frac{1}{z}\right)\\
&+z^{-\beta }\frac{\Gamma (\gamma )  \Gamma (\alpha -\beta ) }{\Gamma (\alpha ) \Gamma (\gamma -\beta )}{}\ _2F_1\left(\beta ,\beta -\gamma +1;\beta-\alpha +1;-\frac{1}{z}\right)
\end{aligned}
\end{equation}
we see that the asymptotic behaviour of $f_L^\lambda(\rho)$ is
\begin{equation}\label{eq:rho_zero}
f^\lambda_L(\rho)=\begin{cases}
\left(\frac{\rho}{m}\right)^L+\cdots & \rho\ll m\\
\frac{(d+2 \Lambda-3)!}{(d+2 \Lambda-4)!!}\frac{(d+2 L-2)!!}{(d+L+\Lambda-2)!}\left(\frac{\rho}{m}\right)^\Lambda+\cdots & \rho\gg m
\end{cases}
\end{equation}

From this expression we see that for $\lambda<d$ the function $f^\lambda_L(\rho)$ grows sub-linearly with $\rho$, while for $\lambda>d$ the growth is super-linear. The edge case $\lambda=d$ is of particular interest: for example, for $\lambda=d$, the lowest modes $L=0$ and $L=1$ reproduce the functions $k^\mu$, because $mf^d_0=\sqrt{\rho^2+m^2}$ (see~\cite[eq.~15.1.8]{AbramowitzStegun}) and $mf^d_1=\rho$:
\begin{equation}
\begin{aligned}
m\chi_{00\cdots0}^d&\sim \sqrt{\boldsymbol k^2+m^2}\\
m\chi_{\ell_1\ell_2\cdots1}^d&\sim \text{linear combinations of}\ k^i
\end{aligned}
\end{equation}
and therefore $k^\mu\in\{\chi_\ell^\lambda\}$, as required. The higher-order modes $L\ge 2$ have a more complicated expression, but they are all 
linear in $\rho$ for large $\rho$. This is an essential property of the radial function $f_L^d(\rho)$, because it guarantees that the integral that defines $ P_\ell^d$ converges as long as that that defines $ P^\mu$ does. We note that if $\lambda<d$, the integral is still convergent, so the admissible range for $\lambda$ is $0\le\lambda\le d$, which is equivalent to $0\le \Lambda\le 1$. If we wish to pick a larger value of $\lambda$, we must impose a faster decay rate for the Fourier modes $a_{\vec k}$ (for arbitrary $\lambda$, we need $|\boldsymbol k|^\Lambda |a|^2\in L^1(\mathbb R^d,\d k)$).

\remark{the second argument of $f_L^\lambda$ in the hypergeometric function~\eqref{eq:radial_f_12} is $\frac12(L-\Lambda)$. Therefore, if $\Lambda$ is an integer, then this argument vanishes for $L=\Lambda$, in which case the radial function becomes $f=\rho^L$. This in turns means that the mode $\chi_\ell^\lambda$, for $L=\Lambda\in\mathbb N$, is a (harmonic) polynomial in $\boldsymbol k$ of order $L$, that is, $\chi\sim \text{linear combinations of}\ k^{i_1}k^{i_2}\cdots k^{i_\Lambda}$.}

Now that we have characterised the modes $\chi_\ell^\lambda$, we can go back to our initial intention of enlarging the Poincar\'e algebra into the $\mathfrak{bms}$ algebra. To this end, let us note that $\mathcal M_{\mu\nu}$ commutes with $(\Delta-\lambda)$, and therefore
\begin{equation}
(\Delta-\lambda)\mathcal M_{\mu\nu}\chi_\ell^\lambda=0
\end{equation}
which implies that $\mathcal M_{\mu\nu}\chi_\ell^\lambda$ must be a linear combination of the modes themselves:
\begin{equation}\label{eq:mode_algebra}
\mathcal M_{\mu\nu}\chi^\lambda_\ell=\sum_{\alpha\in\mathbb Z^{d-1}}c^\alpha_{\mu\nu}(\lambda,\ell)\,\chi^\lambda_{\ell+\alpha}
\end{equation}
where $c^\alpha_{\mu\nu}$ are a set of constants. This is precisely the relation we were after,~\eqref{eq:BMS_algebra}.

\remark{
we have now completed the construction of the modes $\chi$, and showed that they satisfy a closed algebra of the form $[\mathcal M,\chi]=\chi$.
A crucial ingredient of this construction was the fact that  the differential operator $(\Delta-\lambda)$ commutes with $\mathcal M_{\mu\nu}$. In fact, it possible to prove that $(\Delta-\lambda)$ is the most general second order differential operator that commutes with $\mathcal M_{\mu\nu}$, so our construction is in a sense unique: if we were to insist that the modes $\chi$  have to be given by a differential equation, then this equation has to be $(\Delta-\lambda)\chi=0$.
}

The coefficients $c_{\mu\nu}^\alpha(\lambda,\ell)$ are the structure constants of the $\lambda$-extended $\mathfrak{bms}_{d+1}$ algebra:
\begin{equation}\label{eq:lambda_bms_algebra}
\begin{aligned}
\{ P_\ell^\lambda, P_{\ell'}^{\lambda'}\}&=0\\
\{ M_{\alpha \beta},  M^{\mu \nu}\}&=4i\delta^{[\mu}{}_{[\alpha}   M^{\nu]}{}_{\beta]}\\
\{ P^\lambda_\ell, M_{\mu\nu}\}&=\sum_{\alpha\in\mathbb Z^{d-1}}c^\alpha_{\mu\nu}(\lambda,\ell) P^\lambda_{\ell+\alpha}
\end{aligned}
\end{equation}

We note that this algebra is uncountable infinite-dimensional, its elements being labelled by the integers $\ell\in\mathbb Z^{d-1}$ and the real number $\Lambda\in[0,1]$ (or a larger interval if we impose a faster decay rate on the Fourier modes). It contains the standard (gravitational) algebra, corresponding to the value $\lambda=d$, which itself contains the Poincar\'e algebra, corresponding to  all the integer values of $|\ell_1|\le\ell_2\le\dots\ell_{d-1}=1$:
\begin{equation}
\lambda\mathfrak{bms}_{d+1}\supset \mathfrak{bms}_{d+1}\supset \mathfrak{iso}(1,d)
\end{equation}
of dimensions $2^{\aleph_0}> \aleph_0>\frac{1}{2} (d+1) (d+2)$.

The algebra that is relevant to the gravitational problem is the one corresponding to $\lambda=d$, since it contains a Poincar\'e sub-algebra generated by the super-translations with $L=0,1$ (see Fig.~\ref{fig:bms_structure}).

\begin{linespread}{1.0} \selectfont
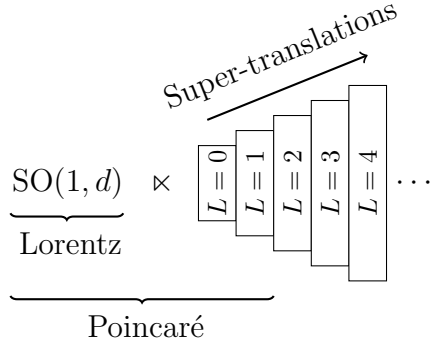
\begin{figure}[!h]
\centering
\begin{tikzpicture}
\node at (.3,0) {$\mathrm{SO}(1,d)\ \ \, \ltimes$};

\begin{scope}[shift={(1,0)}]
\filldraw[fill=white, draw=black] (1.2-.5,-.5) rectangle (1.7-.5,0.5) node[pos=.5, rotate=90] {\footnotesize$L=0$};
\filldraw[fill=white, draw=black] (1.2,-.7) rectangle (1.7,0.7) node[pos=.5, rotate=90] {\footnotesize$L=1$};
\filldraw[fill=white, draw=black] (1.7,-.9) rectangle (2.2,0.9) node[pos=.5, rotate=90] {\footnotesize$L=2$};
\filldraw[fill=white, draw=black] (2.2,-1.1) rectangle (2.7,1.1) node[pos=.5, rotate=90] {\footnotesize$L=3$};
\filldraw[fill=white, draw=black] (2.7,-1.3) rectangle (2.7+.5,1.3) node[pos=.5, rotate=90] {\footnotesize$L=4$};
\node at (3.6,0) {$\cdots$};
\end{scope}

\draw[decoration={brace,mirror},decorate,thick]  (-.8,-.4) -- (.7,-.4);
\node at (0,-.8) [align=center] {Lorentz};

\draw[->,thick] [rotate=22] (2,.1) -- (4.3,.1);

\node at (2.8,1.7) [rotate=22] {Super-translations};

\draw[decoration={brace,mirror},decorate,thick]  (-.8,-1.5) -- (2.7,-1.5);
\node at (1,-1.9) [align=center] {Poincar\'e};

\end{tikzpicture}
\caption{The structure of the $\lambda\mathfrak{bms}_{d+1}$ algebra, for $\lambda=d$. Each box represents the set of super-translations $\{ P^d_\ell\}$ with $|\ell_1|\le\ell_2\le\dots\ell_{d-1}= L\in\mathbb N$.}
\label{fig:bms_structure}
\end{figure}
\end{linespread}

A rather straightforward method to calculate the structure constants $c^\alpha_{\mu\nu}(\lambda,\ell)$ is to use the orthogonality of the spherical harmonics:
\begin{equation}
c_{\mu\nu}^{\alpha}(\lambda,\ell)=\frac{1}{f^\lambda_{\ell_{d-1}+\alpha_{d-1}}(\rho)}\int Y^*_{ \ell+\alpha}(\theta) \mathcal M_{\mu\nu}\chi^\lambda_{\ell}(\rho,\theta)\,\mathrm d\Omega_{d-1}
\end{equation}
which is in fact independent of $\rho$, so we can take $\rho\to 0$ or any other convenient value. In the appendix~\ref{sec:coefficients} we quote the coefficients $c_{\mu\nu}^{\alpha}(\lambda,\ell)$ for arbitrary $\lambda$, for $d=3,4,5$.

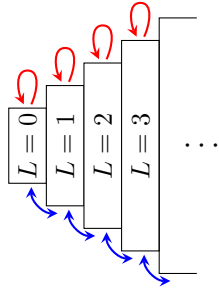
\begin{figure}[!h]
\centering
\begin{tikzpicture}

\draw [thick,red,->,>=stealth] (1.5,.55) to[out=65,in=0] (1.45,1) to[out=180,in=180-65] (1.4,.55);
\draw [thick,red,->,>=stealth] (1.5+.5,.55+.3) to[out=65,in=0] (1.45+.5,1+.3) to[out=180,in=180-65] (1.4+.5,.55+.3);
\draw [thick,red,->,>=stealth] (1.5+1,.55+.6) to[out=65,in=0] (1.45+1,1+.6) to[out=180,in=180-65] (1.4+1,.55+.6);
\draw [thick,red,->,>=stealth] (1.5+1.5,.55+.9) to[out=65,in=0] (1.45+1.5,1+.9) to[out=180,in=180-65] (1.4+1.5,.55+.9);

\filldraw[fill=white, draw=black] (1.2,-.5) rectangle (1.7,0.5) node[pos=.5, rotate=90] {\footnotesize$L=0$};
\filldraw[fill=white, draw=black] (1.7,-.8) rectangle (2.2,0.8) node[pos=.5, rotate=90] {\footnotesize$L=1$};
\filldraw[fill=white, draw=black] (2.2,-1.1) rectangle (2.7,1.1) node[pos=.5, rotate=90] {\footnotesize$L=2$};
\filldraw[fill=white, draw=black] (2.7,-1.4) rectangle (2.7+.5,1.4) node[pos=.5, rotate=90] {\footnotesize$L=3$};
\draw[black] (2.7+.5+.5,-1.7) -- (2.7+.5,-1.7) -- (2.7+.5,1.7) -- (2.7+.5+.5,1.7);
\node at (3.8,0) {$\cdots$};

\draw [thick,blue,<->,>=stealth] (1.7+.15+1,-.9-.6) to[out=180,in=-90] (1.2+.3+1,-.55-.6);
\draw [thick,blue,<->,>=stealth] (1.7+.15+1-1,-.9-.6+.6) to[out=180,in=-90] (1.2+.3+1-1,-.55-.6+.6);
\draw [thick,blue,<->,>=stealth] (1.7+.15+.5,-.9-.3) to[out=180,in=-90] (1.2+.3+.5,-.55-.3);
\draw [thick,blue,<->,>=stealth] (1.7+.15+1.5,-.9-.9) to[out=180,in=-90] (1.2+.3+1.5,-.55-.9);

\end{tikzpicture}
\caption{The inhomogeneous part of $\lambda\mathfrak{bms}_{d+1}$. The red and blue arrows represent the action of rotations and boosts on the generators of super-translation. Rotations do not change the value of $L$, while boosts take us from $L$ to $L\pm1$.
}
\label{fig:bms_lorentz_action}
\end{figure}

As the differential operators $\mathcal M_{\mu\nu}$ have one unit of angular momentum (\`{a} la Wigner-Eckart), the indices $\alpha_i$ take values in $\alpha_i\in\{0,\pm1\}$. One should also note that, as the angular momentum generators only act on the angular part of the modes, the spatial coefficients $c_{ij}^\alpha$ are only non-zero if $\alpha_{d-1}=0$ (so that $L\to L$). On the other hand, the boost operators include a radial derivative $\partial_\rho$, so they do change the value of $L$. This means that $c^\alpha_{0i}$ will be non-zero for $\alpha_{d-1}=\pm1$ (see Fig.~\ref{fig:bms_lorentz_action}).

Another interesting fact about the coefficients $c^\alpha_{\mu\nu}$ is that, when $\alpha_{d-1}=+1$, they are proportional to $(\ell_{d-1}(\ell_{d-1}+d-1)-\lambda)$. Therefore, if $\lambda=\Lambda (\Lambda+d-1)$ with $\Lambda\in\mathbb N$, then $c^\alpha_{\mu\nu}(\lambda,\ell)|^{\alpha_{d-1}=1}_{\ell_{d-1}=\Lambda}$ vanishes, which means that we get a closed sub-algebra, of dimension
\begin{equation}\label{eq:dim_Lambda}
\begin{aligned}
\dim&\,(\mathrm{SO}(1,d))+\sum_{\ell_{d-1}=0}^\Lambda\ \sum_{\ell_{d-2}=0}^{\ell_{d-1}}\cdots\sum_{\ell_2=0}^{\ell_3}\ \sum_{\ell_1=-\ell_2}^{\ell_2}1\\
&=\frac{1}{2} d (d+1)+\frac{(d+\Lambda-2)!}{(d-1)!\, \Lambda!}(d+2 \Lambda-1)
\end{aligned}
\end{equation}

\remark{a simple argument to prove that when $L=\Lambda\in\mathbb N$ the action of the Lorentz operators do not raise the value of $L$ (that is, that the coefficient corresponding to $L\to L+1$ vanishes) is to realise that for $L=\Lambda$, the mode $\chi_\ell^\lambda$ is given by a linear combination of monomials of the form $k^{i_1}k^{i_2}\cdots k^{i_\Lambda}$, and therefore the derivatives $\mathcal M_{0i}=ik_0\partial_i$ cannot raise the power of $\rho=|\boldsymbol k|$ (it can only lower it or leave it as is). Therefore, we cannot have a term with $\chi_{L+1}\sim \rho^{L+1}$ in the expansion of $\mathcal M_{0i}\chi_\ell$.}

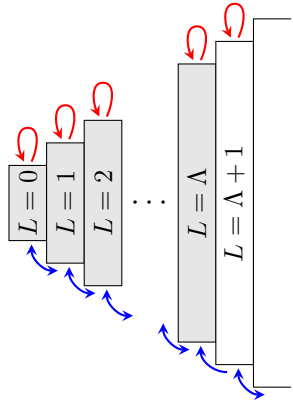
\begin{figure}[!h]
\centering
\begin{tikzpicture}

\draw [thick,red,->,>=stealth] (1.5,.55) to[out=65,in=0] (1.45,1) to[out=180,in=180-65] (1.4,.55);
\draw [thick,red,->,>=stealth] (1.5+.5,.55+.3) to[out=65,in=0] (1.45+.5,1+.3) to[out=180,in=180-65] (1.4+.5,.55+.3);
\draw [thick,red,->,>=stealth] (1.5+1,.55+.6) to[out=65,in=0] (1.45+1,1+.6) to[out=180,in=180-65] (1.4+1,.55+.6);
\draw [thick,red,->,>=stealth] (3.75,1.9) to[out=65,in=0] (3.7,2.35) to[out=180,in=180-65] (3.65,1.9);
\draw [thick,red,->,>=stealth] (3.75+.5,1.9+.3) to[out=65,in=0] (3.7+.5,2.35+.3) to[out=180,in=180-65] (3.65+.5,1.9+.3);

\filldraw[fill=black!10!white, draw=black] (1.2,-.5) rectangle (1.7,0.5) node[pos=.5, rotate=90] {\footnotesize$L=0$};
\filldraw[fill=black!10!white, draw=black] (1.7,-.8) rectangle (2.2,0.8) node[pos=.5, rotate=90] {\footnotesize$L=1$};
\filldraw[fill=black!10!white, draw=black] (2.2,-1.1) rectangle (2.7,1.1) node[pos=.5, rotate=90] {\footnotesize$L=2$};
\node at (3.1,0) {$\cdots$};
\filldraw[fill=black!10!white, draw=black] (3.45,-1.85) rectangle (3.95,1.85) node[pos=.5, rotate=90] {\footnotesize$L=\Lambda$};
\filldraw[fill=white, draw=black] (3.95,-2.15) rectangle (4.45,2.15) node[pos=.5, rotate=90] {\footnotesize$L=\Lambda+1$};
\draw[black] (4.95,-2.45) -- (4.45,-2.45) -- (4.45,2.45) -- (4.95,2.45);

\draw [thick,blue,<->,>=stealth] (1.7+.15+1,-.9-.6) to[out=180,in=-90] (1.2+.3+1,-.55-.6);
\draw [thick,blue,<->,>=stealth] (1.7+.15+1-1,-.9-.6+.6) to[out=180,in=-90] (1.2+.3+1-1,-.55-.6+.6);
\draw [thick,blue,<->,>=stealth] (1.7+.15+.5,-.9-.3) to[out=180,in=-90] (1.2+.3+.5,-.55-.3);
\draw [thick,blue,<->,>=stealth] (1.7+.15+2.25-.5,-.9-1.35+.3) to[out=180,in=-90] (1.2+.3+2.25-.5,-.55-1.35+.3);
\draw [thick,blue,->,>=stealth] (1.7+.15+2.25,-.9-1.35) to[out=180,in=-90] (1.2+.3+2.25,-.55-1.35);
\draw [thick,blue,<->,>=stealth] (1.7+.15+2.25+.5,-.9-1.35-.3) to[out=180,in=-90] (1.2+.3+2.25+.5,-.55-1.35-.3);

\end{tikzpicture}
\caption{The result of Fig.~\ref{fig:bms_lorentz_action} when $\Lambda$ is an integer. The grey boxes represents the closed sub-algebra. If $\Lambda=1$, this sub-algebra is Poincar\'e.}
\label{fig:bms_j}
\end{figure}

In the particular case $\lambda=d$, that is, $\Lambda=1$, the second term of~\eqref{eq:dim_Lambda} becomes $d+1$, which is the number of space-time translations in a manifold of $d$ spatial dimensions. This is the Poincar\'e sub-algebra. For $\Lambda=0$ we get $\mathfrak{so}(1,d)\times\mathfrak{u}(1)$. For other integer values of $\Lambda$, we get other closed sub-algebras (see Fig.~\ref{fig:bms_j}), whose relevance and physical interpretation is not clear yet. For non-integer values of $\Lambda$, there are no closed sub-algebras.

\begin{figure}[h!]
\hspace{-24pt}
\begin{tikzpicture}
\node at (0,0) {\includegraphics[width=.37\textwidth]{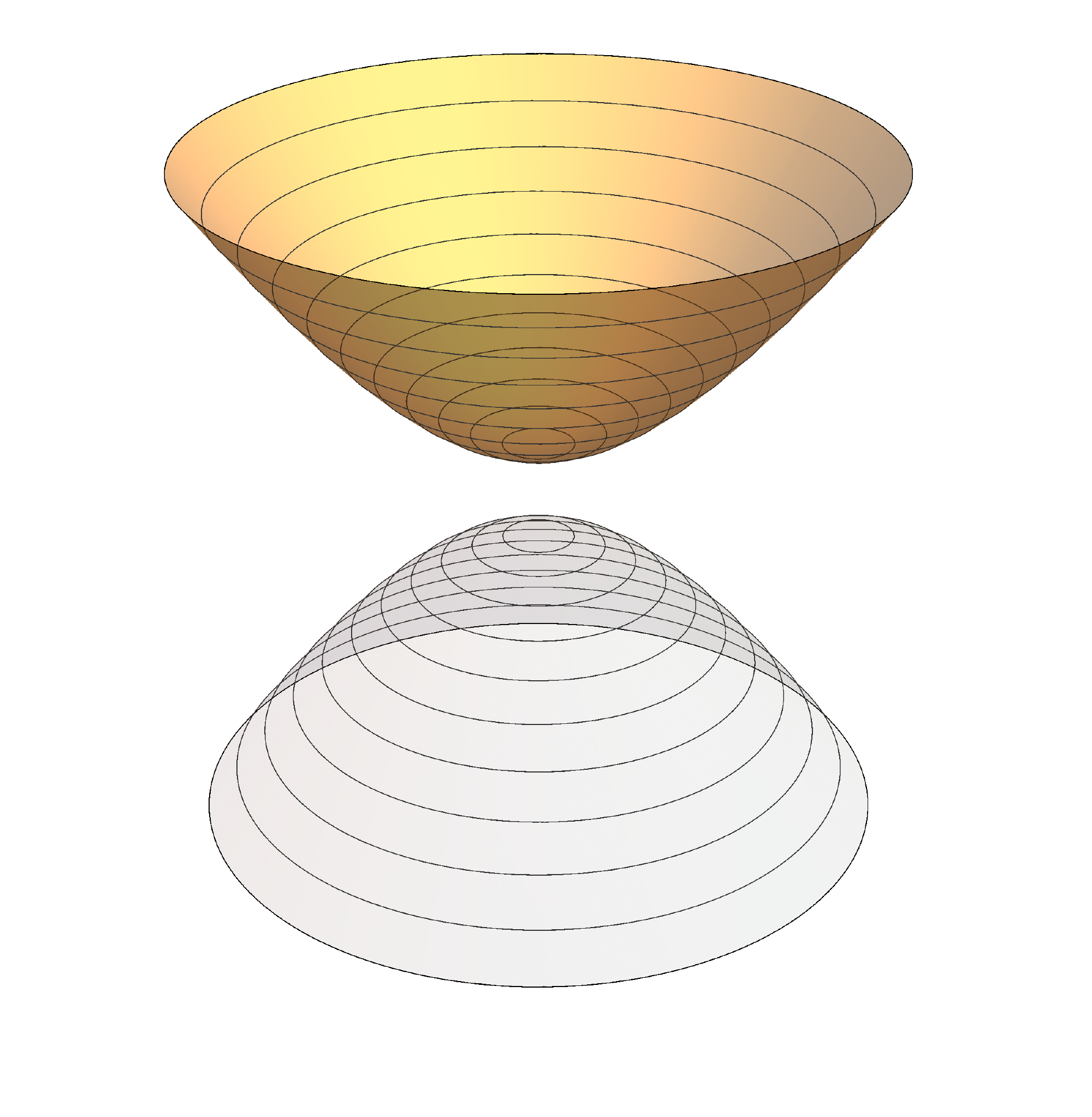}};
\node at (-5.5,0) {\includegraphics[width=.37\textwidth]{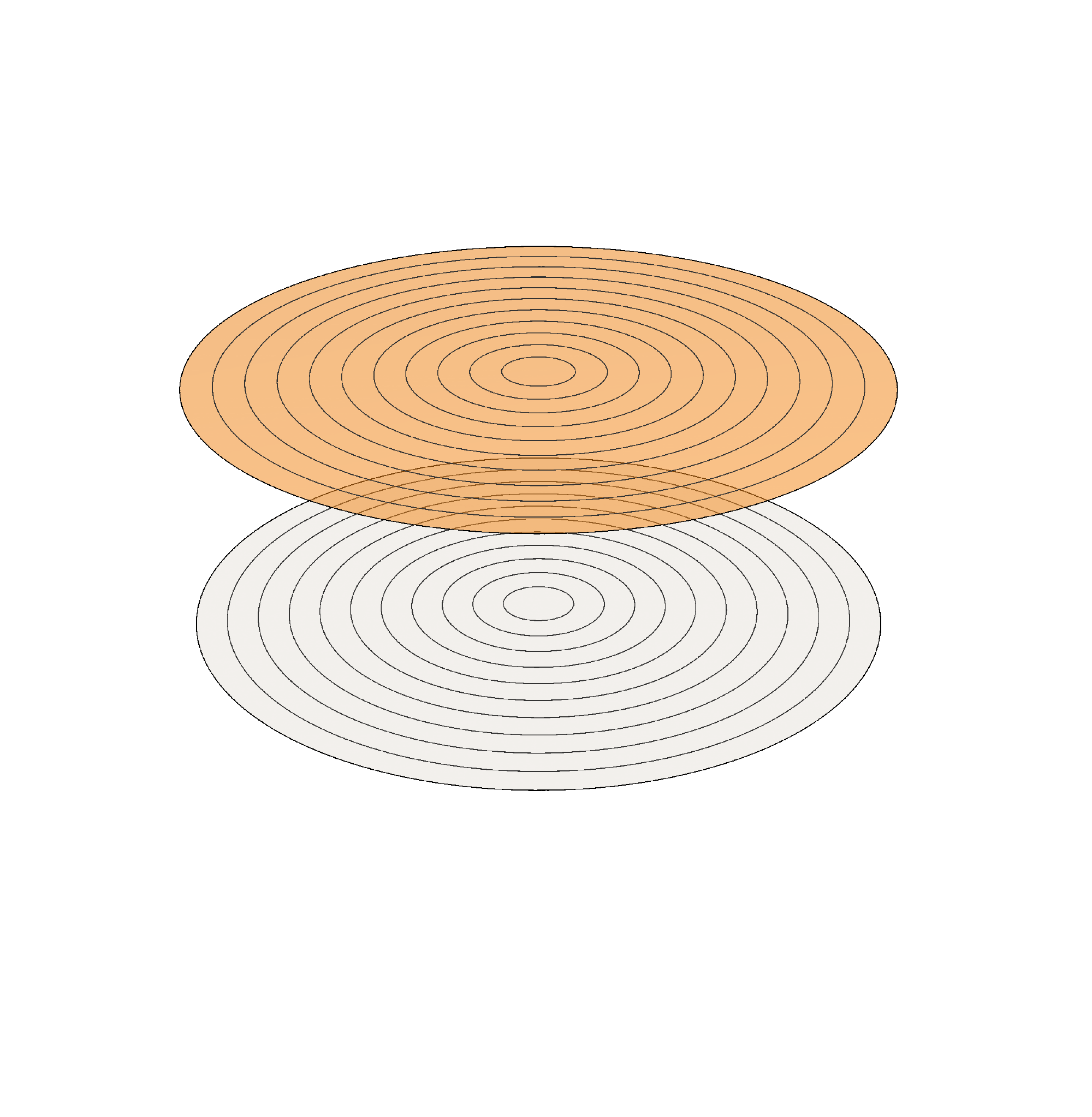}};
\node at (5.5,0) {\includegraphics[width=.37\textwidth]{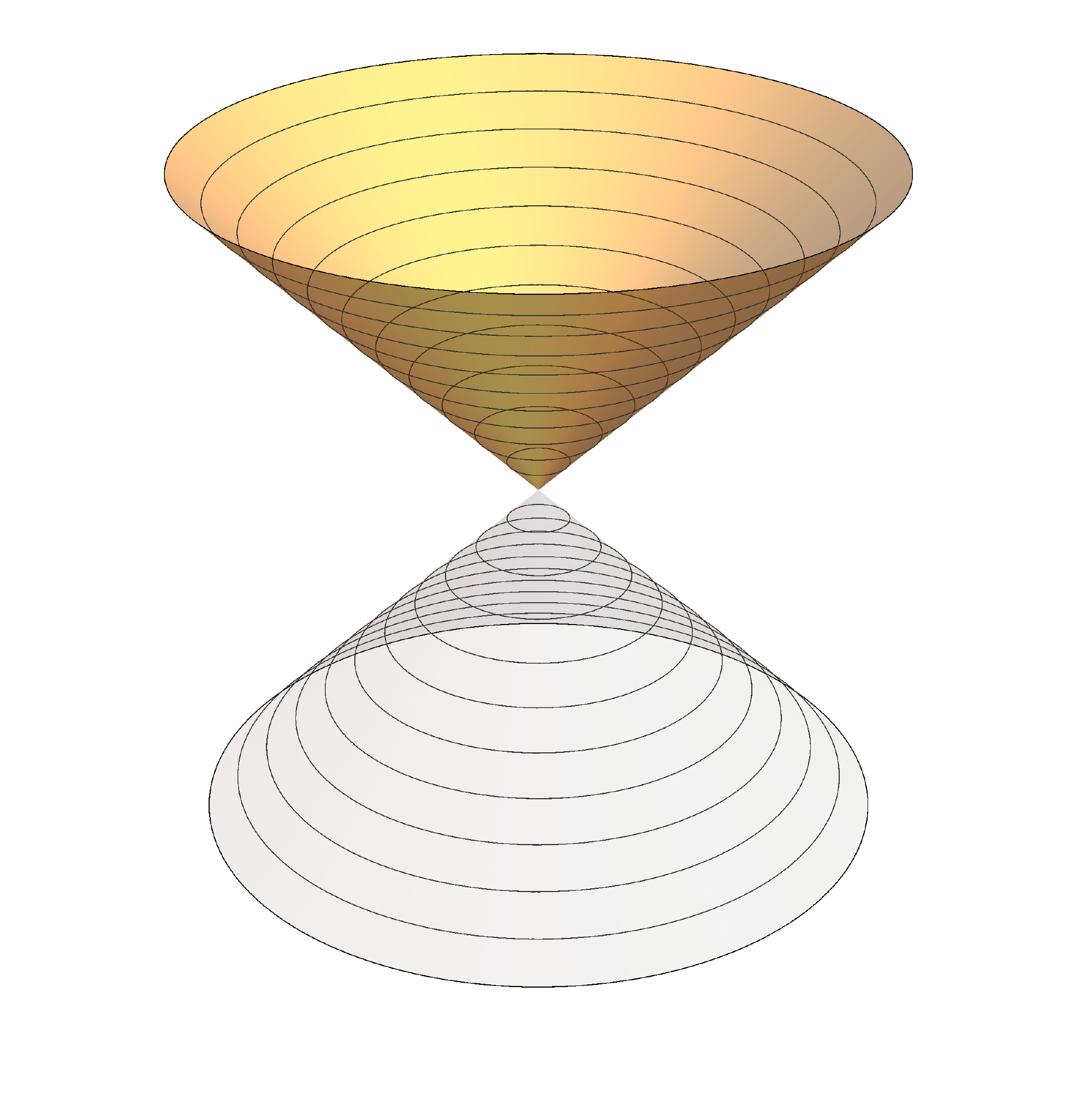}};
\draw[->,thick,>=stealth] (-1.8,.4) -- (-3.3,.4);
\draw[->,thick,>=stealth] (1.8,.4) -- (3.3,.4);
\node at (-2.5,.7) {\small$m\to\infty$};
\node at (2.55,.7) {\small$m\to0$};
\end{tikzpicture}
\vspace{-34pt}
\caption{The mass hyperboloid $k^2+m^2=0$, and its non-relativistic (left) and massless (right) limits, as given by $\rho\ll m$ and $\rho\gg m$ respectively.}
\label{fig:mass_hyp}
\end{figure}
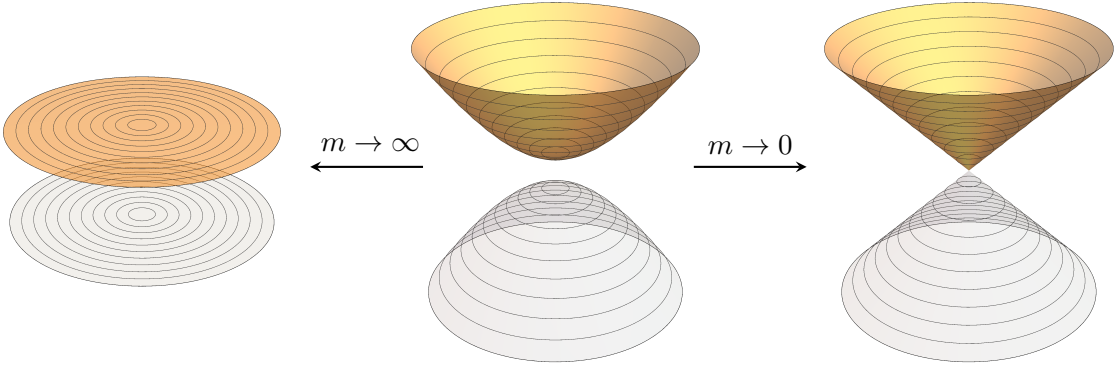

In any case, equation~\eqref{eq:lambda_bms_algebra} proves that our construction does indeed lead to an infinite-dimensional generalisation of the Poincar\'e algebra. We will next discuss how this construction is related to the gravitational $\mathfrak{bms}$ algebra. The correspondence between the canonical $\mathfrak{bms}$ algebra and the gravitational one is most clear in what we will call the massless limit of the former, to be discussed below. Let us therefore discuss this limit, together with the opposite limit, the non-relativistic limit. These two limits correspond to taking $\rho\gg m$ and $\rho\ll m$, respectively (see Fig.~\ref{fig:mass_hyp}).

Let us begin with the massless limit  $m\to 0$. In this limit, the mass hyperboloid becomes singular, and the Laplacian blows up. One may nevertheless expand the Laplacian, the zero-modes, and the $\mathfrak{bms}$ algebra in power series in $m$, and keep the leading terms. The result is
\begin{equation}
\begin{aligned}
\tilde \Delta&\eqdef \lim_{m\to 0}\Delta=\rho^2\partial_\rho^2+d\,\rho\partial_\rho\\
\tilde{\mathcal M}_{0i}&\eqdef \lim_{m\to 0}\mathcal M_{0i}=i\rho\partial_i\\
\tilde{\mathcal M}_{ij}&\eqdef \lim_{m\to 0}\mathcal M_{ij}=2ik_{[i}\partial_{j]}\\
\tilde \chi_\ell^{\lambda}&\eqdef\lim_{m\to 0}m^\Lambda\chi_\ell^\lambda=\frac{(d+2 \Lambda-3)!}{(d+2 \Lambda-4)!!}\frac{(d+2 L-2)!!}{(d+L+\Lambda-2)!}\ \rho^\Lambda\ Y_\ell(\theta)
\end{aligned}
\end{equation}
where we have used the limit $\rho\gg m$ as given by~\eqref{eq:rho_zero}. From here on, a tilde $\tilde\cdot$ over an object means that it corresponds to the massless limit of the same.

These modes satisfy the same algebra as the standard modes, with the exact same coefficients, $\tilde c^\alpha_{\mu\nu}=c^\alpha_{\mu\nu}$. As the radial function $f^\lambda_L(\rho)$ is a power function instead of a hypergeometric one, the calculation of the structure constants $c^\alpha_{\mu\nu}(\lambda,\ell)$ is particularly simple in this limit. 

If we take the particular case $\lambda=d$, the massless modes become
\begin{equation}
\tilde \chi_\ell^d=\frac{(d-1)!}{(d-2)!!}\frac{(d+2 L-2)!!}{(d+L-1)!}\ \rho\ Y_\ell(\theta),
\end{equation}
to be compared with the asymptotic Killing fields, $\xi(\mathscr I^+)$, as given by~\eqref{eq:killing_d}. The $L$-dependent prefactor accounts for the different normalisation between the canonical and gravitational normalisations.

We can now consider the opposite limit, the non-relativistic limit. In the limit $m\to\infty$, the mass hyperboloid becomes flat, and the Laplace operator becomes
\begin{equation}
\hat\Delta\eqdef m^2\lim_{m\to\infty}\frac{1}{m^2}\Delta=m^2\partial_i^2
\end{equation}
which is, once again the Casimir operator, but now of the homogeneous Galilei Group instead of the Lorentz Group. The Galilei Group is generated by
\begin{equation}\label{eq:NR_generators}
\begin{aligned}
\hat{\mathcal M}_{0i}&\eqdef m\lim_{m\to 0}\frac{1}{m}\mathcal M_{0i}=im\partial_i\\
\hat{\mathcal M}_{ij}&\eqdef \lim_{m\to 0}\mathcal M_{ij}=2ik_{[i}\partial_{j]}
\end{aligned}
\end{equation}
such that $\hat \Delta=\hat{\mathcal M}_{0i}\hat{\mathcal M}^{0i}$. Here and in the remainder of this document, we will place a hat $\hat\cdot$ over an object to indicate that it corresponds to the non-relativistic counterpart of the corresponding relativistic object.

In this limit, the modes $\hat\chi_\ell^\lambda$ become
\begin{equation}\label{eq:nr_modes_limit}
\begin{aligned}
\hat\chi_\ell(\rho,\theta)&\eqdef\frac{1}{m^L}\lim_{m\to\infty}m^L\chi_\ell^\lambda\\
&=\left(\frac{\rho}{m}\right)^LY_\ell(\theta)
\end{aligned}
\end{equation}
as given by the $\rho\ll m$ limit of our previous $f^\lambda_L(\rho)$, cf.~\eqref{eq:rho_zero}. In this limit, the modes become independent of $\lambda$. This was to be expected, because
\begin{equation}\label{eq:nr_limit_delta}
\frac{1}{m^2} \Delta\chi^\lambda_\ell=\frac{\lambda}{m^2}\chi^\lambda_\ell\xrightarrow{\: m\to\infty \: }0,
\end{equation}
which is independent of $\lambda$, and therefore so are non-relativistic modes (which satisfy $\hat\Delta\hat\chi=0$, being the non-relativistic limit of $\Delta\chi=\lambda\chi$). This is consistent with the fact that the functions $k^i$ satisfy $\hat \Delta k^i=0$, where $\hat \Delta$ is the flat Laplacian. Therefore, our modes $\hat\chi_\ell$ contain the functions $k^i$ for $L=1$ (the case $L=0$ plays the role of the central charge of the Bargmann group, $\hat \chi_0=\text{const.}$, as we shall discuss in the following chapter).

Unlike in the massless case, and due to the factor of $m$ in the non-relativistic limit of the boost generators (cf.~\eqref{eq:NR_generators}), here the algebra is not the same as in the relativistic case. If we write
\begin{equation}
\hat{\mathcal M}_{\mu\nu} \hat\chi_\ell=\sum_{\alpha} \hat c_{\mu\nu}^{\alpha}(\ell)\hat\chi_{\ell+\alpha}
\end{equation}
then the new structure constants $\hat c_{\mu\nu}$ are given by
\begin{equation}\label{eq:nrbms_can_limit}
\begin{aligned}
\hat c^\alpha_{ij}&=c^\alpha_{ij}\\
\hat c^\alpha_{0i}&=\begin{cases}
c^\alpha_{0i} & \alpha_{d-1}=-1\\
0 & \alpha_{d-1}=+1
\end{cases}
\end{aligned}
\end{equation}

We call this new algebra the $\mathfrak{nrbms}$ algebra, as it corresponds to the non-relativistic limit of the canonical realisation of the $\mathfrak{bms}$ algebra. For consistency, one should realise this non-relativistic algebra in terms of a scalar field with a Galilean symmetry instead of a relativistic field. We postpone this construction to the next chapter. There we will also consider the non-relativistic limit in more abstract terms, that is, as an \.In\"on\"u-Wigner contraction of the abstract $\mathfrak{bms}$ algebra.

\chapter{Non-relativistic $\mathrm{BMS}$.}\label{sec:nrbms}

In this chapter we will address the problem of a non-relativistic $\mathrm{BMS}$ Group. To define this group, we could try to repeat the analysis of Chapter~\ref{sec:bms_grav} but in the context of Newtonian space-times, characterised by a contravariant degenerate spatial metric $h^{\mu\nu}$ and 
covariant vector $\tau_\mu$; see for example \cite{PhysRevD.31.1841} and references therein. We will not do this here; instead, we will construct a possible candidate for the $\mathfrak{nrbms}$ algebra by performing an \.In\"on\"u-Wigner contraction of the relativistic $\mathfrak{bms}$ algebra (in the same spirit as the Bargmann algebra can be obtained by contracting the Poincar\'e one~\cite{1953PNAS...39..510I}); and we will then provide an explicit realisation of the contracted algebra, by means of a free field with Galilean space-time symmetries, mimicking the construction of Chapter~\ref{sec:bms_can}. This chapter is essentially a generalisation of the results of~\cite{2017arXiv170503739B} to an arbitrary number of space-time dimensions.

\section{The algebra $\mathfrak{nrbms}$ as a contraction of $\mathfrak{bms}$.}
\label{section2}
The $\mathfrak{bms}$ algebra is the semi-direct sum of the Lorentz algebra with the generators of the super-translations, which form an infinite-dimensional abelian sub-algebra. In a similar fashion, the $\mathfrak{nrbms}$ algebra will be given by the semi-direct sum of the Bargmann algebra with the generators of super-translations.

The canonical realisation of the $\lambda$-extended $\mathfrak{bms}_{d+1}$ algebra in terms of the Fourier modes of  relativistic free field leads to the  algebra~\eqref{eq:lambda_bms_algebra}
\begin{equation}\label{eq:lambda_bms_algebra_2}
\begin{aligned}
\{ P_\ell^\lambda, P_{\ell'}^{\lambda'}\}&=0\\
\{ M_{\alpha \beta},  M^{\mu \nu}\}&=4i\delta^{[\mu}{}_{[\alpha}   M^{\nu]}{}_{\beta]}\\
\{ P^\lambda_\ell, M_{\mu\nu}\}&=\sum c^\alpha_{\mu\nu}(\lambda,\ell) P^\lambda_{\ell+\alpha}\\
\end{aligned}
\end{equation}
where $M_{\mu \nu}$  are the  generators of Lorentz transformations, and $P_\ell^\lambda$ are the generators of super-translations. To simplify the discussion, we shall henceforth set $\lambda\equiv d$, and omit this label altogether.  As discussed in the previous chapter, for this value of $\lambda$ the algebra above contains a Poincar\'e sub-algebra. This sub-algebra is spanned by $M_{\mu\nu},P_\ell$, with $\ell_{d-1}=0,1$. These operators form a closed sub-algebra of $\mathfrak{bms}_{d+1}$ because the structure constants $c^\alpha_{\mu\nu}(\ell)$ vanish when $\alpha_{d-1}=+1$ and $\ell_{d-1}\ge 1$.

In the particular case $d=3$, the algebra above agrees with the algebra of the gravitational $\mathfrak{bms}$~\eqref{bmsalgebra_grav} (modulo an $\ell_2$-dependent prefactor, resulting from the different normalisation for the generators of super-translations). We have argued that the correspondence should hold for any number of space dimensions $d$.

We now proceed to perform the contraction of this algebra in order to obtain its non-relativistic analogue. In order to accommodate the central extension of the Galilei algebra, we consider the direct product of the $\mathrm{BMS}$ Group with $U(1)$, with generator $ Z$, and introduce the following transformation: 
\begin{equation}\label{eq:omega_contrac}
\begin{split}
H&\eqdef\omega(P_0+ Z)\\
\hat Z&\eqdef\frac{1}{\omega}(P_0- Z)\\
\hat M_{0i}&\eqdef\frac{1}{\omega}M_{0i}\\
\hat M_{ij}&\eqdef M_{ij}
\end{split}\qquad\overset{\text{inverse}}{\Longleftrightarrow}\qquad
\begin{split}
P_0&=\frac{1}{2\omega}H+\omega \hat Z\\
Z&=\frac{1}{2\omega}H-\omega \hat Z\\
M_{0i}&=\omega \hat M_{0i}\\
M_{ij}&=\hat M_{ij}
\end{split}
\end{equation}
together with
\begin{equation}\label{eq:super_generators_contrac}
\hat P_\ell\eqdef\omega^{f(L)}P_\ell\qquad\overset{\text{inverse}}{\Longleftrightarrow}\qquad P_\ell=\omega^{-f(L)}\hat P_\ell, \quad L\geq 1,
\end{equation}
where $L=\ell_{d-1}$ and $f(L)$ is an unspecified function, and $\omega$ is a dimensionless parameter which we shall take $\omega\to\infty$ at the end.

In the limit $\omega\to\infty$, the centrally-extended relativistic algebra~\eqref{eq:lambda_bms_algebra_2} becomes
\begin{equation}\label{nrbmsalgebra}
\begin{split}
{}[\hat M_{ij},\hat M_{mn}]&=4i\delta_{[i[m}\hat J_{n]j]}\\
[\hat M_{ij},\hat M_{0m}]&=2i\delta_{m[i}\hat M_{j]0}\\
[\hat M_{0i},\hat M_{0j}]&=0
\end{split}\hspace{40pt}
\begin{split}
[\hat P_\ell,H]&=0\\
[\hat M_{ij},H]&=0\\
[\hat M_{0i},H]&=i\hat P_i
\end{split}\hspace{40pt}
\begin{split}
[\hat P_\ell,\hat P_{\ell'}]&=0\\
[\hat M_{ij},\hat P_\ell]&=\sum\hat c^\alpha_{ij} \hat P_{\ell+\alpha}\\
[\hat M_{0i},\hat P_\ell]&=\sum\hat c^\alpha_{0i} \hat P_{\ell+\alpha}
\end{split}
\end{equation}
where $\hat c^\alpha_{ij}\eqdef c^\alpha_{ij}$ and
\begin{equation}
\hat c^\alpha_{0i}\eqdef\lim_{\omega\to\infty}\omega^{f(L)-f(L')-1}\ c^\alpha_{0i}
\end{equation}

In order to have a non-trivial $\omega\to\infty$ limit, we must have
\begin{equation}
f(L)-f(L')-1\equiv0
\end{equation}
for some $L'$. Moreover, the structure constants $c^\alpha_{0i}$ are non-zero only if $|L'-L|=1$. This means that the only non-trivial contractions are the ones that satisfy
\begin{equation}
f(L)=f(0)\pm L
\end{equation}

Furthermore, in order to obtain the Bargmann algebra as a sub-algebra,  for $L=1$ we should recover the standard contraction $\hat P_i=P_i$, so that $f(1)=0$. With this,
\begin{equation}\label{eq:two_signs}
f(L)=\pm(L-1)
\end{equation}

The conclusion of this discussion is that if we restrict ourselves to scalings of the form $\hat P_\ell=\omega^{f(L)}P_\ell$ then there are only two non-trivial contractions of the $\mathfrak{bms}_{d+1}$ algebra, corresponding to either sign in $f(L)=\pm(L-1)$, which we will call $\mathfrak{nrbms}^{(\downarrow)}$ (corresponding to the plus sign) and $\mathfrak{nrbms}^{(\uparrow)}$ (corresponding to the negative sign). In either case, the structure constants are given by
\begin{equation}\label{nrmatrices}
\begin{split}
&\ \underline{\ f^{(\downarrow)}(L)=L-1\ }\\
\hat c^\alpha_{ij}&=c^\alpha_{ij}\\
\hat c^\alpha_{0i}&=\begin{cases}c^\alpha_{0i} & \alpha_{d-1}=-1\\ 0 &  \alpha_{d-1}=+1\end{cases}
\end{split}\qquad\qquad\qquad
\begin{split}
&\ \underline{\ f^{(\uparrow)}(L)=1-L\ }\\
\hat c^\alpha_{ij}&=c^\alpha_{ij}\\
\hat c^\alpha_{0i}&=\begin{cases}0 & \alpha_{d-1}=-1\\c^\alpha_{0i} & \alpha_{d-1}=+1\end{cases}
\end{split}
\end{equation}

\remark{the first of these algebras is the exact same algebra we obtained by the non-relativistic limit $\rho\ll m$ of the modes $\chi_\ell$ (cf.~\eqref{eq:nrbms_can_limit}). This was actually to be expected, because the factors of $\omega$ in the abstract contraction~\eqref{eq:omega_contrac},~\eqref{eq:super_generators_contrac} are the same as the factors of $m$ in the explicit limit~\eqref{eq:NR_generators},~\eqref{eq:nr_modes_limit}. The limits $\omega\to\infty$ and $m\to\infty$ are formally identical. The second algebra is new.}

The two possible algebras above contain time and space translations, rotations, boosts, super-translations, and a central charge, and they both contain a Bargmann sub-algebra (see~Fig.\ref{fig:nrbms_algebra}).

\vspace{-15pt}

\begin{linespread}{1.0} \selectfont
\begin{figure}[!h]
\centering
\begin{tikzpicture}
\node at (-1.45,0) {$\mathrm{SO}(d)\ \ltimes\ \mathbb R^d\ \, \ltimes\,$};

\begin{scope}[shift={(1.6,0)}]
\filldraw[fill=white, draw=black] (1.2-.5,-.5) rectangle (1.7-.5,0.5) node[pos=.5, rotate=90] {\footnotesize$L=1$};
\filldraw[fill=white, draw=black] (1.2,-.7) rectangle (1.7,0.7) node[pos=.5, rotate=90] {\footnotesize$L=2$};
\filldraw[fill=white, draw=black] (1.7,-.9) rectangle (2.2,0.9) node[pos=.5, rotate=90] {\footnotesize$L=3$};
\filldraw[fill=white, draw=black] (2.2,-1.1) rectangle (2.7,1.1) node[pos=.5, rotate=90] {\footnotesize$L=4$};
\filldraw[fill=white, draw=black] (2.7,-1.3) rectangle (2.7+.5,1.3) node[pos=.5, rotate=90] {\footnotesize$L=5$};
\node at (3.6,0) {$\cdots$};
\end{scope}

\begin{scope}[shift={(.6,0)}]
\draw[->,thick] [rotate=22] (2,.1) -- (4.3,.1);

\node at (2.8,1.7) [rotate=22] {Super-translations};
\end{scope}

\draw[decoration={brace},decorate,thick]  (-2.95,.4) -- (-.55,.4);
\node at (-1.75,1) [align=center] {Homogeneous\\Galilei};
\node at (1.15,0) {$\mathbb R\ \times\ \mathbb R\ \,\times\,$};

\draw[decoration={brace,mirror},decorate,thick]  (.18,-.4) -- (1.55,-.4);

\node at (.85,-1) [align=center] {Central charge\\Time translations};

\draw[decoration={brace,mirror},decorate,thick]  (-2.95,-1.6) -- (2.8,-1.6);
\node at (0,-2.1) [align=center] {Bargmann};

\end{tikzpicture}
\caption{The structure of the $\mathfrak{nrbms}^{(\updownarrow)}$ algebras. Each box represents an abelian algebra corresponding to the super-translations $\{\hat P_\ell\}$, with fixed $L=\ell_{d-1}$.
}
\label{fig:nrbms_algebra}
\end{figure}
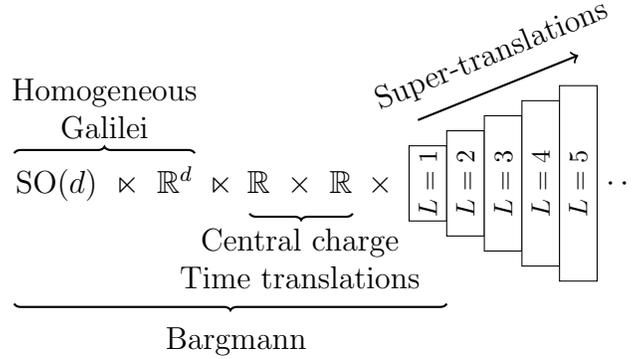
\end{linespread}

It is important to note that in the case of the $\mathfrak{nrbms}^{(\downarrow)}$ contraction, the boost operators lower the value of $L$ to $L-1$. This is in stark contrast with the relativistic case, where we have simultaneous contributions from $L-1$ and $L+1$. Therefore, unlike in the relativistic case, here the algebra resulting from the `$\downarrow$' contraction contains an infinite number of finite-dimensional sub-algebras, obtained by considering all the super-translation generators with $1\le L\le \Lambda$ for given $\Lambda\in\mathbb N$; the dimension of these sub-algebras is given by~\eqref{eq:dim_Lambda} (plus one, due to the central charge). All these sub-algebras contain a Bargmann sub-algebra, and the associated matrices
$\hat c^\alpha_{ij},\hat c^\alpha_{0i}$ provide an infinite number of finite-dimensional representations of the homogeneous Galilei group (finite-dimensional indecomposable representations  of homegeneous Galilei have been studied in \cite{0305-4470-39-29-026}).

On the other hand, in the case of the $\mathfrak{nrbms}^{(\uparrow)}$ algebra, the boost operators raise the value of $L$ to $L+1$, which means that the only  finite-dimensional sub-algebra of $\mathfrak{nrbms}^{(\uparrow)}$ is the Bargmann algebra (see~Fig.\ref{fig:nrbms_algebra_pm}).

\begin{linespread}{1.0} \selectfont
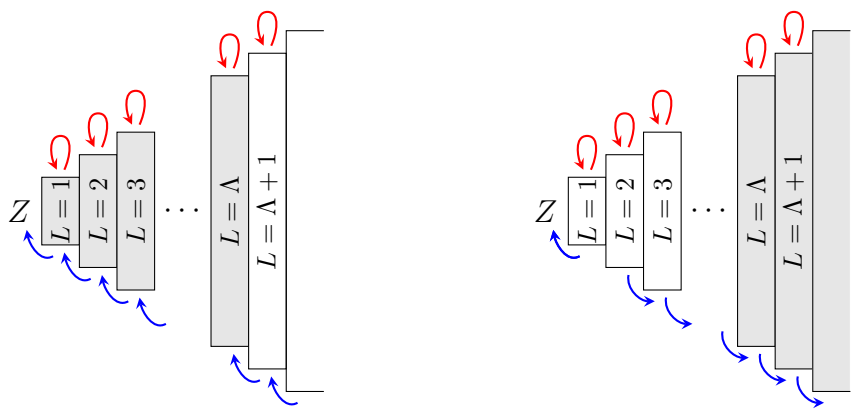
\begin{figure}[!h]
\centering
\begin{tikzpicture}

\node at (.9,0) {$Z$};

\draw [thick,red,->,>=stealth] (1.5,.55) to[out=65,in=0] (1.45,1) to[out=180,in=180-65] (1.4,.55);
\draw [thick,red,->,>=stealth] (1.5+.5,.55+.3) to[out=65,in=0] (1.45+.5,1+.3) to[out=180,in=180-65] (1.4+.5,.55+.3);
\draw [thick,red,->,>=stealth] (1.5+1,.55+.6) to[out=65,in=0] (1.45+1,1+.6) to[out=180,in=180-65] (1.4+1,.55+.6);
\draw [thick,red,->,>=stealth] (3.75,1.9) to[out=65,in=0] (3.7,2.35) to[out=180,in=180-65] (3.65,1.9);
\draw [thick,red,->,>=stealth] (3.75+.5,1.9+.3) to[out=65,in=0] (3.7+.5,2.35+.3) to[out=180,in=180-65] (3.65+.5,1.9+.3);

\filldraw[fill=black!10!white, draw=black] (1.2,-.45) rectangle (1.7,0.45) node[pos=.5, rotate=90] {\footnotesize$L=1$};
\filldraw[fill=black!10!white, draw=black] (1.7,-.75) rectangle (2.2,0.75) node[pos=.5, rotate=90] {\footnotesize$L=2$};
\filldraw[fill=black!10!white, draw=black] (2.2,-1.05) rectangle (2.7,1.05) node[pos=.5, rotate=90] {\footnotesize$L=3$};
\node at (3.1,0) {$\cdots$};
\filldraw[fill=black!10!white, draw=black] (3.45,-1.8) rectangle (3.95,1.8) node[pos=.5, rotate=90] {\footnotesize$L=\Lambda$};
\filldraw[fill=white, draw=black] (3.95,-2.1) rectangle (4.45,2.1) node[pos=.5, rotate=90] {\footnotesize$L=\Lambda+1$};
\draw[black] (4.95,-2.4) -- (4.45,-2.4) -- (4.45,2.4) -- (4.95,2.4);

\draw [thick,blue,->,>=stealth] (1.7+.15-.5,-.9+.3) to[out=180+30,in=-90+10] (1.2+.3-.5,-.55+.3);
\draw [thick,blue,->,>=stealth] (1.7+.15,-.9) to[out=180+30,in=-90+10] (1.2+.3,-.55);
\draw [thick,blue,->,>=stealth] (1.7+.15+.5,-.9-.3) to[out=180+30,in=-90+10] (1.2+.3+.5,-.55-.3);
\draw [thick,blue,->,>=stealth] (1.7+.15+1,-.9-.6) to[out=180+30,in=-90+10] (1.2+.3+1,-.55-.6);

\draw [thick,blue,->,>=stealth] (1.7+.15+2.25,-.9-1.35) to[out=180+30,in=-90+10] (1.2+.3+2.25,-.55-1.35);
\draw [thick,blue,->,>=stealth] (1.7+.15+2.25+.5,-.9-1.35-.3) to[out=180+30,in=-90+10] (1.2+.3+2.25+.5,-.55-1.35-.3);

\begin{scope}[shift={(7,0)}]
\node at (.9,0) {$Z$};

\draw [thick,red,->,>=stealth] (1.5,.55) to[out=65,in=0] (1.45,1) to[out=180,in=180-65] (1.4,.55);
\draw [thick,red,->,>=stealth] (1.5+.5,.55+.3) to[out=65,in=0] (1.45+.5,1+.3) to[out=180,in=180-65] (1.4+.5,.55+.3);
\draw [thick,red,->,>=stealth] (1.5+1,.55+.6) to[out=65,in=0] (1.45+1,1+.6) to[out=180,in=180-65] (1.4+1,.55+.6);
\draw [thick,red,->,>=stealth] (3.75,1.9) to[out=65,in=0] (3.7,2.35) to[out=180,in=180-65] (3.65,1.9);
\draw [thick,red,->,>=stealth] (3.75+.5,1.9+.3) to[out=65,in=0] (3.7+.5,2.35+.3) to[out=180,in=180-65] (3.65+.5,1.9+.3);

\filldraw[fill=white, draw=black] (1.2,-.45) rectangle (1.7,0.45) node[pos=.5, rotate=90] {\footnotesize$L=1$};
\filldraw[fill=white, draw=black] (1.7,-.75) rectangle (2.2,0.75) node[pos=.5, rotate=90] {\footnotesize$L=2$};
\filldraw[fill=white, draw=black] (2.2,-1.05) rectangle (2.7,1.05) node[pos=.5, rotate=90] {\footnotesize$L=3$};
\node at (3.1,0) {$\cdots$};
\filldraw[fill=black!10!white, draw=black] (3.45,-1.8) rectangle (3.95,1.8) node[pos=.5, rotate=90] {\footnotesize$L=\Lambda$};
\filldraw[fill=black!10!white, draw=black] (3.95,-2.1) rectangle (4.45,2.1) node[pos=.5, rotate=90] {\footnotesize$L=\Lambda+1$};
\filldraw[fill=black!10!white,draw=white] (4.45,-2.4) rectangle (4.95,2.4);
\draw[black] (4.95,-2.4) -- (4.45,-2.4) -- (4.45,2.4) -- (4.95,2.4);

\draw [thick,blue,->,>=stealth] (1.7+.15-.5,-.9+.3) to[out=180+30,in=-90+10] (1.2+.3-.5,-.55+.3);

\draw [thick,blue,->,>=stealth] (1.7+.15-.5,-.9+.3) to[out=180+30,in=-90+10] (1.2+.3-.5,-.55+.3);
\draw [thick,blue,<-,>=stealth] (1.7+.15+1,-.9-.6) to[out=180,in=-90-10] (1.2+.3+1,-.55-.6);
\draw [thick,blue,<-,>=stealth] (1.7+.15+.5,-.9-.3) to[out=180,in=-90-10] (1.2+.3+.5,-.55-.3);
\draw [thick,blue,<-,>=stealth] (1.7+.15+2.25-.5,-.9-1.35+.3) to[out=180,in=-90-10] (1.2+.3+2.25-.5,-.55-1.35+.3);
\draw [thick,blue,<-,>=stealth] (1.7+.15+2.25,-.9-1.35) to[out=180,in=-90-10] (1.2+.3+2.25,-.55-1.35);
\draw [thick,blue,<-,>=stealth] (1.7+.15+2.25+.5,-.9-1.35-.3) to[out=180,in=-90-10] (1.2+.3+2.25+.5,-.55-1.35-.3);

\end{scope}

\end{tikzpicture}
\caption{The inhomogeneous part of $\mathfrak{nrbms}^{(\updownarrow)}$. The red and blue arrows represent the action of rotations and boosts on the generators of super-translation. Rotations do not change the value of $L$, while boosts take us from $L$ to $L\pm1$. The grey boxes represent the different sub-algebras, obtained from varying $\Lambda\in\mathbb N$ (in the first case, they are all finite-dimensional, while in the second case they are infinite-dimensional).}
\label{fig:nrbms_algebra_pm}
\end{figure}
\end{linespread}

In the following section we will construct an explicit realisation of the $\mathfrak{nrbms}^{(\downarrow)}_{d+1}$ algebra corresponding to the plus sign contraction, and we will also discuss the possibility of adding dilatations and expansions.

\section{Canonical realisation of the $\mathfrak{nrbms}$ algebra.} 
\label{section3}

We now proceed to construct  the canonical realisation of the $\mathfrak{nrbms}^{(\downarrow)}_{d+1}$ algebra. The construction will be analogous to the canonical realisation of $\mathfrak{bms}_{d+1}$ from the previous chapter: we want to study the symmetries of galilean equations of motion of a free fields. As before, we will frame our discussion directly in terms of the Fourier modes of the non-relativistic field, $a,a^*$, in terms of which one may realise the Bargmann algebra through
\begin{equation}\label{eq:non-relativistic_noether}
\begin{aligned}
\hat Z&\eqdef\int a^*(\boldsymbol k)m\, a(\boldsymbol k)\,\frac{\mathrm d\boldsymbol k}{(2\pi)^d}\\
H&\eqdef\int a^*(\boldsymbol k)\frac{\boldsymbol k^2}{2m} a(\boldsymbol k)\,\frac{\mathrm d\boldsymbol k}{(2\pi)^d}\\
\hat P_i&\eqdef\int a^*(\boldsymbol k)k_i a(\boldsymbol k)\,\frac{\mathrm d\boldsymbol k}{(2\pi)^d}\\
\hat M_{0i}&\eqdef t\hat P_i+\int a^*(\boldsymbol k)\hat{\mathcal M}_{0i} a(\boldsymbol k)\,\frac{\mathrm d\boldsymbol k}{(2\pi)^d}\\
\hat M_{ij}&\eqdef\int a^*_{\vec k} \hat{\mathcal M}_{ij} a_{\vec k}\,\frac{\mathrm d\boldsymbol k}{(2\pi)^d}
\end{aligned}
\end{equation}
where $\hat Z$ is the central charge which generates $\mathrm U(1)$ rotations (here, and due to a lack of Coleman-Mandula~\cite{1967PhRv..159.1251C}, external symmetries and internal symmetries mix in a non-trivial way). The differential operators $\hat{\mathcal M}_{0i},\hat{\mathcal M}_{ij}$ are given by~\eqref{eq:NR_generators}:
\begin{equation}
\begin{aligned}
\hat{\mathcal M}_{0i}&=im\partial_i\\
\hat{\mathcal M}_{ij}&=2ik_{[i}\partial_{j]}
\end{aligned}
\end{equation}
and they satisfy the (homogeneous) Galilei algebra
\begin{equation}
\begin{aligned}
[\hat{\mathcal M}_{ij},\hat{\mathcal M}_{mn}]&=4i\delta_{[i[m}\hat{\mathcal M}_{n]j]}\\
[\hat{\mathcal M}_{ij},\hat{\mathcal M}_{0m}]&=2i\delta_{m[i}\hat{\mathcal M}_{j]0}\\
[\hat{\mathcal M}_{0i},\hat{\mathcal M}_{0j}]&=0
\end{aligned}
\end{equation}


On the other hand, the Noether charges satisfy the the Bargmann algebra,
\begin{equation}\label{eq:barg_algebra}
\begin{split}
\{\hat M_{ij},\hat M_{mn}\}&=4i\delta_{[i[m}\hat M_{n]j]}\\
\{\hat M_{ij},\hat M_{0m}\}&=2i\delta_{m[i}\hat M_{j]0}\\
\{\hat M_{0i},\hat M_{0j}\}&=0
\end{split}\hspace{30pt}
\begin{split}
\{\hat P_\ell,H\}&=0\\
\{\hat M_{ij},H\}&=0\\
\{\hat M_{0i},H\}&=i\hat P_i
\end{split}\hspace{30pt}
\begin{split}
\{\hat P_i,\hat P_j\}&=0\\
\{\hat M_{ij},\hat P_m\}&=2i\delta_{m[i}\hat P_{j]}\\
\{\hat M_{0i},\hat P_m\}&=i\delta_{im}\hat Z
\end{split}
\end{equation}

What we want to do is to generalise the last column of this algebra, by introducing a set of functions $\hat\chi_\ell$ in place of the standard momentum,
\begin{equation}
\begin{aligned}
k^i&\to m\hat \chi_\ell(\vec k)\\
\hat P^i&\to \hat P_\ell\eqdef m\int a^*_{\vec k}\hat \chi_\ell(\vec k) a_{\vec k}\,\frac{\mathrm d\boldsymbol k}{(2\pi)^d}
\end{aligned}
\end{equation}
such that $k_i\in\{\hat\chi_\ell\}$ for some values of $\ell$, thus extending the Bargmann algebra into the $\mathfrak{nrbms}$ algebra.

As before, let us consider the quadratic Casimir operator of the homogeneous Galilei group, 
\begin{equation}
\hat \Delta\eqdef\hat{\mathcal M}_{0i}\hat{\mathcal M}^{0i}=m^2\partial_i^2
\end{equation}
which coincides with the Laplace-Beltrami operator on $\mathbb R^d$ (flat space). With this, we will define our modes as the zero-modes of $\hat\Delta$:
\begin{equation}
\hat \Delta\hat \chi_\ell(\boldsymbol k)=0
\end{equation}
together with the same angular dependence as in the relativistic case (cf.~\eqref{eq:mode_angular_eig}). As a consistency check, we note that, once again, the momentum $k^i$ satisfies $\hat \Delta k^i=0$, so the family of zero-modes $\{\hat\chi\}$ will contain the functions $k^i$ as a subset.

The general solution to the equation above is
\begin{equation}
\hat \chi_\ell(\rho,\theta)=f_L(\rho)Y_\ell(\theta)
\end{equation}
where $\rho,\theta$ are the spherical components of $k_i$, and $Y_\ell$ are the spherical harmonics. On the other hand, $f_L$ is given by the solution to
\begin{equation}
f''+\frac{d-1}{z}f'-\frac{L(L+d-2)}{z^2}f=0
\end{equation}
where $z=\rho/m$, and a prime denotes differentiation with respect to $z$. The solution to this equation is
\begin{equation}
f_L(r)=c_1\left(\frac{\rho}{m}\right)^L+c_2\left(\frac{m}{\rho}\right)^{L+d-2}
\end{equation}

As before, we set $c_2\equiv0$ to avoid the singular behaviour at $\rho\to 0$, and set $c_1\equiv 1$. With this,
\begin{equation}\label{eq:non-relativistic_omega}
\hat \chi_\ell(\rho,\theta)=\left(\frac{\rho}{m}\right)^L Y_\ell(\theta)
\end{equation}

We see that the $L=0$ mode corresponds to the central charge (being momentum-independent), while the $L=1$ modes agree with the spherical components of $\boldsymbol k$, so that, as expected, the family $\{\hat \chi_\ell\}$ contains the functions $k_i$ as a special subcase. Comparing~\eqref{eq:non-relativistic_omega} with the equation~\eqref{eq:nr_modes_limit}, we see that the non-relativistic modes agree with the non-relativistic limit of the relativistic modes from the previous chapter, as the notation suggests. Therefore, they satisfy the $\mathfrak{nrbms}^{(\downarrow)}_{d+1}$ algebra
\begin{equation}\label{eq:nrmodes_diff}
\hat{\mathcal M}_{\mu\nu} \hat\chi_\ell=\sum_{\alpha} \hat c_{\mu\nu}^{\alpha}(\ell)\hat\chi_{\ell+\alpha}
\end{equation}
where the structure constants are given by~\eqref{eq:nrbms_can_limit}:
\begin{equation}
\begin{aligned}
\hat c^\alpha_{ij}&=c^\alpha_{ij}\\
\hat c^\alpha_{0i}&=\begin{cases}
c^\alpha_{0i} & \alpha_{d-1}=-1\\
0 & \alpha_{d-1}=+1
\end{cases}
\end{aligned}
\end{equation}

With this, we define the generators of super-translations  as
\begin{equation}\label{nrPl}
\hat P_\ell\eqdef m\int a^*_{\boldsymbol k}\hat \chi_\ell(\boldsymbol k) a_{\boldsymbol k}\,\frac{\mathrm d\boldsymbol k}{(2\pi)^d}
\end{equation}

One should note that, unlike the relativistic case, here the existence of $\hat P_\ell$ is not guaranteed by the existence of $\hat P_i$, because the non-relativistic modes $\hat \chi$ scale as $\rho^L$ for large $\rho$ instead of linearly with $\rho$. Therefore, $a_{\boldsymbol k}$ being square-integrable is not enough for the integral defining $\hat P_\ell$ to converge; we must impose the stronger condition that $|\boldsymbol k|^L |a_{\boldsymbol k}|^2$ is integrable for all $L\in\mathbb N$. If we only require for a finite number of super-translations to exist, those corresponding to $0\le L\le\Lambda$ for a certain integer $\Lambda$, then we must impose that $|\boldsymbol k|^L |a_{\boldsymbol k}|^2$ is integrable for all $0\le L\le\Lambda$. This doesn't interfere with the closedness of the sub-algebra, because the structure constants $\hat c_{\mu\nu}^\alpha$ only connect super-translations of order $L$ among themselves and with super-translations of order $L-1$ (see the left diagram of Fig.~\ref{fig:nrbms_algebra_pm}).

In any case, using~\eqref{eq:nrmodes_diff} we see that the functions $\hat P_\ell$ satisfy the algebra
\begin{equation}
\{\hat P_\ell,\hat M_{\mu\nu}\}=\sum_{\alpha\in\mathbb Z^{d-1}}\hat c^\alpha_{\mu\nu}(\ell) \hat P_{\ell+\alpha}\\
\end{equation}
which, together with the first two columns of~\eqref{eq:barg_algebra}, agrees with the $\mathfrak{nrbms}^{(\downarrow)}_{d+1}$ algebra~\eqref{nrmatrices}. In principle, it may possible to construct a realisation of $\mathfrak{nrbms}_4^{(\uparrow)}$ by using the second solution to the radial equation, $f_L\sim \rho^{2-L-d}$, but the fact that these functions are singular at the origin implies that the Fourier modes $a_{\boldsymbol k}$ must go to zero faster than any polynomial if we want the integral that defines $\hat P_\ell$ to converge. We will not consider this possibility any further here.

\remark{we have chosen the definition of $\hat\chi$ to be $\hat\Delta\hat\chi\equiv 0$ in order to match the non-relativistic limit of the relativistic modes (cf.~\eqref{eq:nr_limit_delta}). We could have worked with a more general equation $(\hat\Delta-\hat\lambda)\hat \chi=0$ for an arbitrary real constant $\hat\lambda>0$. The solutions $\{\hat\chi^{\hat\lambda}_\ell\}$ satisfy a more general algebra than $\mathfrak{nrbms}^{(\downarrow)}$, but such that it reduces to the it when we take $\hat\lambda\equiv 0$. In order to obtain this generalised algebra by contracting the $\lambda\mathfrak{bms}$ algebra, one must take the $\omega,\lambda\to\infty$ limit of the latter, while keeping $\hat\lambda\equiv \lambda/\omega$ fixed. We state without proof that the radial equation for general $\hat\lambda$ reads $f^{\hat\lambda}_L(z)=z^L \, _0F_1\left(;\frac{d}{2}+L;-\frac{\hat\lambda}{4} z^2  \right)$ where $_0F_1$ is the confluent hypergeometric function, and that the structure constants for general $\hat\lambda$ are $\hat c^\alpha_{\mu\nu}(\hat\lambda,\ell)=c^\alpha_{\mu\nu}(\hat \lambda+\ell_{d-1}(\ell_{d-1}+d-1),\ell)$. We will not study this extension of $\mathfrak{nrbms}$ here. In what follows, we will restrict ourselves to the case $\hat\lambda\equiv0$, that is, the regular $\mathfrak{nrbms}^{(\downarrow)}$ algebra, where the modes are given by $\hat \chi_\ell(\boldsymbol k)=\left(\frac{\rho}{m}\right)^LY_\ell(\theta)$.}

The algebra $\mathfrak{nrbms}_{d+1}^{(\downarrow)}$ has a very important property, one that is not present in the relativistic $\lambda\mathfrak{bms}_{d+1}$ case: here, the modes $\hat\chi(\boldsymbol k)$ are actually homogeneous polynomials in $k^i$, of degree $L$. This means that the symmetries generated by $\hat P_\ell$ are \emph{local} when acting on the field $\phi$, meaning that 
\begin{equation}
\delta_\ell\phi(x)=\{\hat P_\ell,\phi(x)\}=\hat\chi_\ell(-i\partial)\phi(x)
\end{equation}
where $\hat\chi_\ell$ is a harmonic polynomial of degree $L$. In other words, $\hat \chi_\ell(-i\partial)$ is nothing but a polynomial in $\partial_i$:
\begin{equation}
\hat \chi_\ell(-i\partial)=\sum_{|\alpha|=L}\mathcal C^\alpha\partial_\alpha
\end{equation}
for a certain set of coefficients $\mathcal C^\alpha$. Differential operators of this form (and generalisations thereof), in the context of symmetries of partial differential equations, have been studied extensively in the literature; see for example \cite{Nikitin1991,boyer1976,2009arXiv0912.0789V}.

Furthermore, the fact that the polynomials $\hat\chi_\ell(\boldsymbol k)$ are \emph{homogeneous} and of degree $L$ implies that they satisfy
\begin{equation}
\mathcal D\hat\chi_\ell=L\hat\chi_\ell
\end{equation}
where $\mathcal D$ is the homogeneity operator, $\mathcal D\eqdef k^i\partial_i=\rho\partial_\rho$. This means that if we define the dilatation operator as
\begin{equation}
D\eqdef 2tH+i\int a^*_{\boldsymbol k} \mathcal D a_{\boldsymbol k}\,\frac{\mathrm d\boldsymbol k}{(2\pi)^d}
\end{equation}
then the generators of super-translations satisfy
\begin{equation}
\{D,\hat P_\ell\}=iL \hat P_\ell
\end{equation}
which extends the $\mathfrak{nrbms}$ algebra to include dilatations, giving rise to a ``Weyl-$\mathfrak{nrbms}$''. We note that the main obstruction of this extension to the relativistic algebra is the presence of a non-zero mass $m$, which introduces a length scale to the theory (so that it is not invariant under dilatations). From a more pragmatic point of view, the energy $k_0=\sqrt{\boldsymbol k^2+m^2}$ is not a homogeneous polynomial, so it is not an eigenvector of $\mathcal D$ (unless we take $m\equiv 0$).

Once we realise that the algebra admits dilatations, it becomes natural to ask ourselves about the action of the expansion operator (or Schr\"odinger conformal transformations), given by
\begin{equation}\label{eq:conformal}
C\eqdef-t^2H+tD+\frac m2\int a^*_{\boldsymbol k} \hat \Delta a_{\boldsymbol k}\,\frac{\mathrm d\boldsymbol k}{(2\pi)^d}
\end{equation}
and which satisfies~\cite{1972PhRvD...5..377H}
\begin{equation}
\{D,C\}=2iC,\quad \{H,C\}=iD,\quad \{C,\hat P_i\}=i\hat M_{0i},\ \text{etc.}
\end{equation}

Using
\begin{equation}
\begin{aligned}
{}[\hat \Delta,\hat \chi_\ell]&=\hat\Delta\hat \chi_\ell+2(\partial_i\hat \chi_\ell)\frac{\partial}{\partial k^i}\\
&=\frac{2}{im}\left[\sum_{\alpha} \hat c_{0i}^{\alpha}\,\hat\chi_{\ell+\alpha}\right]\frac{\partial}{\partial k^i}
\end{aligned}
\end{equation}
we obtain
\begin{equation}\label{expansionalgebra}
\{C,\hat P_\ell\}= itL \hat P_\ell-im\int a^*_{\boldsymbol k}\left[\sum_{\alpha} \hat c_{0i}^{\alpha}\,\hat\chi_{\ell+\alpha}\right]\frac{\partial}{\partial k^i} a_{\boldsymbol k}\,\frac{\mathrm d\boldsymbol k}{(2\pi)^d}
\end{equation}

The \emph{r.h.s.} of~\eqref{expansionalgebra} is not an element of $\mathfrak{nrbms}^{(\downarrow)}_{d+1}$. This means that if we attempt to extend the algebra to include $C$, the resulting algebra is not closed, which seems to preclude a possible ``Schr\"odinger-$\mathfrak{nrbms}$''. 

In any case, the \emph{r.h.s.}~\eqref{expansionalgebra} is the bracket of two conserved quantities, which means that it is conserved as well. Indeed, a straightforward calculation confirms that
\begin{equation}
\left[\frac{\partial}{\partial t}+\{\cdot,H\}\right]\{C,\hat P_\ell\}=0
\end{equation}

This means that $\{C,\hat P_\ell\}$, despite not being an element of $\mathfrak{nrbms}$, is a conserved operator, i.e., it generates a symmetry of the equations of motion. In the particular case $L=1$, this commutator agrees with the generators of boosts, $\{C,\hat P_i\}=i\hat M_{0i}$. It is tempting to let $\{C,\hat P_\ell\}$ define a new kind of generator of symmetries, which would generalise the standard generators of boosts; we could dub these objects \emph{super-boosts}. It will be interesting to explore their relation with the relativistic super-rotations.

\chapter*{Conclusions and Outlook.}\addcontentsline{toc}{chapter}{Conclusions and Outlook.}

This completes our study of the $\mathrm{BMS}$ Group. Here we summarise the essential points and highlight those topics that require a further analysis.

\subsubsection*{The BMS Group.}

In Chapter~\ref{sec:bms_grav} we reviewed this group in its original context, the problem of characterising the asymptotic isometries of asymptotically flat space-times. We saw that, while in the strictly flat case the isometries are generated by a finite number of vector fields, once we relax the notion of isometry into an asymptotic isometry, the number of generators becomes infinite. In the first case, these vector fields generate the Poincar\'e Group, and in the second case they generate the $\mathrm{BMS}$ Group, which contains Poincar\'e as a sub-group. Moreover, we saw that one may further extend the symmetry algebra by including the infinite-dimensional set of super-rotations.

Finally, we mentioned that the generalisation of this procedure to higher dimensional manifolds is non-trivial, and there are several opposing views in the literature. The very existence of super-translations depends on the particular falloff rate for the component of the metric, and it is not clear how strong these falloffs should be in arbitrary dimensions. Even more so in the case of super-rotations, whose existence depends on whether the conformal Killing equation on the sphere admits or not an infinite number of solutions; and, as we have mentioned, that only happens in $d=2,3$. Therefore, unless we alter the definition of super-rotations in higher dimensions, one would not expect to find them in $d>3$.

\subsubsection*{Canonical realisation of BMS.}

In Chapter~\ref{sec:bms_can} we generalised the canonical construction of the $\mathfrak{bms}_4$ algebra to an arbitrary number of space-time dimensions, and we argued that this result suggests a non-trivial gravitational $\mathfrak{bms}_{d+1}$ for all $d$. We also generalised this algebra to include an additional label, $\lambda$, which makes the resulting algebra uncountable infinite-dimensional. For some particular values of $\lambda$, the corresponding algebra contains finite-dimensional sub-algebras, such as Poincar\'e for $\lambda=d$.

We have not addressed the problem of super-rotations in the canonical construction. One may presume that, like in the gravitational case, these symmetries are only present in $d=2,3$. These dimensions require a case-by-case analysis.

Another unexplored aspect of the canonical construction is the fact that it admits the extension in terms of the parameter $\lambda$. The particular value $\lambda=d$ was relevant to the gravitational problem, but it would be nice to find a physical system whose group of symmetries is $\lambda\mathfrak{bms}$, with arbitrary $\lambda$.

Finally, we would like to mention that in massless theories the group of external symmetries usually gets enhanced to the Conformal Group, which includes conformal transformations and dilatations together with the standard Poincar\'e transformations. In our canonical construction, it may be possible to do the same in the case of massless super-translations, that is, to consider a conformal $\mathrm{BMS}$, perhaps related to the gravitational conformal $\mathrm{BMS}$ constructed in~\cite{2017arXiv170108110H}.

\subsubsection*{Non-relativistic BMS.}

In Chapter~\ref{sec:nrbms} we studied some possible analogues of $\mathfrak{bms}$ for non-relativistic systems, both by means of a contraction of the abstract relativistic algebra and by the canonical method. We saw that, in a sense, there are only two admissible contractions, and that one of them explicitly arises in the study of the symmetries of Galilei-invariant equations of motion for free fields.

Unlike in the relativistic case, the $\mathfrak{nrbms}$ super-translations are local: they act as polynomials in the differential operator $\partial_i$. Being homogeneous, these polynomials have a simple behaviour under the action of the dilatation operator, which means that the $\mathfrak{nrbms}$ algebra admits an extension that includes dilatations. However, it does not seem to admit conformal transformations, which means that we cannot use the non-relativistic super-translations to extend the Schr\"odinger algebra. The existence of super-rotations in the non-relativistic setting might prove essential in the construction of a possible non-relativistic conformal $\mathrm{NRBMS}$.

Finally, it will be interesting to study the asymptotic symmetries of asymptotically flat Newtonian space-times and to check whether the algebra coincides with the one of the two $\mathfrak{nrbms}$ algebras we have constructed. In other words, it will be nice to have an explicit verification of the correspondence between the canonical $\mathrm{NRBMS}$ Group and the gravitational one, provided the latter actually exists.

\clearpage
\appendix

\chapter{Spherical harmonics.} \label{sec:spherical_harmonics}

Here we quote the form of the spherical harmonics on $S^{d-1}$. A much more detailed discussion can be found in~\cite{2013arXiv1304.2585D,book:Muller,book:Vilenkin}.

The metric on the sphere can be defined recursively through
\begin{equation}
\begin{aligned}
\mathrm ds_1^2&=\mathrm d\theta_1^2\\
\mathrm ds_n^2&=\mathrm d\theta_n^2+\sin^2\theta_n\mathrm ds_{n-1}^2
\end{aligned}
\end{equation}
where $\theta_1\in[0,2\pi)$ and $\theta_n\in[0,\pi)$. The volume element is
\begin{equation}
\begin{aligned}
\mathrm d\Omega_1&=\mathrm d\theta_1\\
\mathrm d\Omega_n&
=\sin^{n-1}\theta_n\ \mathrm d\theta_n\,\mathrm d\Omega_{n-1}
\end{aligned}
\end{equation}

On the other hand, the Laplacian reads
\begin{equation}
\begin{aligned}
\Delta_{S^1}&=\frac{\partial^2}{\partial\theta_1^2}\\
\Delta_{S^n}&=\frac{1}{\sin^{n-1}\theta_n}\frac{\partial}{\partial\theta_n}\left[\sin^{n-1}\theta_n\frac{\partial}{\partial\theta_n}\right]+\frac{1}{\sin^2\theta_n}\Delta_{S^{n-1}}
\end{aligned}
\end{equation}

We define the spherical harmonics through
\begin{equation}
\begin{aligned}
\left[\frac{\partial}{\partial\theta_1}-i\ell_1\right]Y_{\ell_1\ell_2\cdots\ell_{d-1}}(\theta_1,\theta_2,\dots,\theta_{d-1})&=0\\
\left[\Delta_{S^n}+\ell_n(\ell_n+n-1)\right]Y_{\ell_1\ell_2\cdots\ell_{d-1}}(\theta_1,\theta_2,\dots,\theta_{d-1})&=0
\end{aligned}
\end{equation}
where $n=1,2,\dots,d-1$.

The first equation is solved by
\begin{equation}
Y_{\ell_1\ell_2\cdots\ell_{d-1}}(\theta_1,\theta_2,\dots,\theta_{d-1})\propto \mathrm e^{i\theta_1\ell_1}
\end{equation}
where the constant of proportionality is an arbitrary function of $\theta_2,\dots,\theta_{d-1}$. Univaluedness of $Y$ implies that $\ell_1\in\mathbb Z$ is an integer. The rest of equations are solved by
\begin{equation}
Y_{\ell_1\ell_2\cdots\ell_n}(\theta_1,\theta_2,\dots,\theta_n)=Y_{\ell_1\ell_2\cdots\ell_{n-1}}(\theta_1,\theta_2,\dots,\theta_{n-1})\mathscr J(\theta_n)
\end{equation}
where $\mathscr J(\theta)$ satisfies
\begin{equation}
\begin{aligned}
\bigg[\frac{1}{\sin^{n-1}\theta}\frac{\partial}{\partial\theta}\left[\sin^{n-1}\theta\frac{\partial}{\partial\theta}\right]-\frac{\ell_{n-1}(\ell_{n-1}+n-2)}{\sin^2\theta}
+\ell_n(\ell_n+n-1)\bigg]\mathscr J(\theta)=0
\end{aligned}
\end{equation}

If we set $\mathscr J(\theta)=\sin^{\ell_{n-1}}\theta\ y(\cos\theta)$, and make the change of variables $x\equiv\cos\theta$, this equation becomes
\begin{equation}
(1-x^2) y''(x)-(2 \mu+1)x y'(x)+\nu (\nu+2 \mu) y(x)=0
\end{equation}
where $\mu\eqdef \ell_{n-1}+(n-1)/2$ and $\nu\eqdef \ell_n-\ell_{n-1}$. The solution to this differential equation is
\begin{equation}
y(x)= c_1 C_\nu^{(\mu)}(x)+c_2 \left(1-x^2\right)^{\frac{1}{4} (1-2 \mu)} Q_{\nu+\mu-\frac{1}{2}}^{\frac{1}{2}-\mu}(x)
\end{equation}
where $C_\nu^{(\mu)},Q_\nu^\mu$ are the Gegenbauer function and the Legendre function of the second kind~\cite[eqs.~15.4.5, 8.1.3]{AbramowitzStegun}: 
\begin{equation}
\begin{aligned}
C_\nu^{(\mu)}(x)&\eqdef\frac{\Gamma(\nu+2\mu)}{\Gamma(2\mu)\Gamma(\nu+1)}{}_2F_1\left(-\nu,\nu+2\mu;\mu+\frac12;\frac{1-z}{2}\right)\\
Q_\nu^\mu(x)&\eqdef \mathrm e^{i\mu\pi}\frac{\pi^{1/2}\Gamma(\mu+\nu+1)(x^2-1)^{\mu/2}}{2^{\nu+1}x^{\mu+\nu+1}}\ \cdot\\
&\hspace{30pt}\cdot{}_2\tilde F_1\left(\frac{\mu+\nu+2}{2},\frac{\mu+\nu+1}{2};\nu+\frac32;x^{-2}\right)
\end{aligned}
\end{equation}
where $\tilde F$ denotes the regularised hypergeometric function. 

Finiteness at $x=\pm1$ requires $c_2\equiv 0$, and $\nu$ to be a non-negative integer. By induction, $\ell_n\in\mathbb N$, with $|\ell_1|\le\ell_2\le\ell_3\dots\le\ell_n$. The series for $C_\mu^{(\mu)}$ now terminates; the Gegenbauer functions become polynomials:
\begin{equation}
C_\nu^{(\mu)}(x)= \sum_{j=0}^{\lfloor \nu/2\rfloor}(-1)^j\frac{\Gamma(\nu+\mu+j)}{\Gamma(\mu)j!(\nu-2j)!}(2x)^{\nu-2j}
\end{equation}

With this,
\begin{equation}
y_{\ell_n,\ell_{n-1}}^{(n)}(x)\propto C_{\ell_n-\ell_{n-1}}^{\left(\ell_{n-1}+(n-1)/2\right)}(x)\propto \left(\frac{\mathrm d}{\mathrm dx}\right)^{\ell_{n-1}}C_{\ell_n}^{\left((n-1)/2\right)}(x)
\end{equation}

If we normalise the functions to
\begin{equation}
\int_{-1}^{+1} (1-x^2)^{\frac{n}{2}+\ell_{n-1}-1}\left[y_{\ell_n,\ell_{n-1}}^{(n)}(x)\right]^2\,\mathrm dx=1
\end{equation}
we get
\begin{equation}
\begin{aligned}
y_{\ell_n,\ell_{n-1}}^{(n)}(x)&\eqdef(-1)^{\ell_{n-1}} 2^{\ell_{n-1}+\frac{n}{2}-1} \Gamma\! \left(\tfrac{n-1}{2}+\ell_{n-1}\right)\ \cdot\\
&\hspace{25pt}\cdot \sqrt{\tfrac{\left(\ell_n+\frac{n-1}{2}\right) \Gamma (\ell_n-\ell_{n-1}+1)}{\pi\  \Gamma (\ell_n+\ell_{n-1}+n-1)}}\ C_{\ell_n-\ell_{n-1}}^{\left(\ell_{n-1}+(n-1)/2\right)}(x)
\end{aligned}
\end{equation}
where we have included the conventional Condon-Shortley phase $(-1)^{\ell_{n-1}}$.

With this, our spherical harmonics are given by the recursive formula
\begin{equation}
\begin{aligned}
Y_{\ell_1}(\theta_1)&=\frac{1}{\sqrt{2\pi}}\ \mathrm e^{i\theta_1\ell_1}\\
Y_{\ell_1\ell_2\dots\ell_n}(\theta_1,\theta_2,\cdots,\theta_n)&=\sin^{\ell_{n-1}}\!\theta_n\ y_{\ell_n,\ell_{n-1}}^{(n)}(\cos\theta_n)Y_{\ell_1\ell_2\cdots\ell_{n-1}}(\theta_1,\theta_2,\dots,\theta_{n-1})
\end{aligned}
\end{equation}
and they have been normalised to
\begin{equation}
\int Y^*_{\ell_1\ell_2\cdots\ell_n}(\theta)Y_{\ell'_1\ell'_2\cdots\ell'_n}(\theta)\,\mathrm d\Omega_n=\delta_{\ell_1\ell'_1}\delta_{\ell_2\ell'_2}\cdots\delta_{\ell_n\ell'_n}
\end{equation}

\clearpage
\newgeometry{left=2cm,right=2cm}
\chapter{Structure constants.}\label{sec:coefficients}

In this appendix we quote the value of the structure constants $c^\alpha_{\mu \nu}(\lambda,\ell)$ for $d=3,4,5$.

For $d=3$,
\begin{equation*}
\begin{aligned}
c_{01}^{\pm+}&=\frac{1}{2} (\ell_2 (\ell_2+2)-\lambda )\sqrt{\frac{1}{(2 \ell_2+1) (2 \ell_2+3)^3}(\ell_2\pm\ell_1+1) (\ell_2\pm\ell_1+2)}\\
c_{01}^{\pm-}&=\frac{1}{2} \sqrt{\frac{2 \ell_2+1}{2 \ell_2-1} (\ell_2\mp\ell_1) (\ell_2\mp\ell_1-1)}\\
c_{02}^{\pm+}&=\pm\frac{i}{2}  (\ell_2 (\ell_2+2)-\lambda )\sqrt{\frac{1}{(2 \ell_2+1) (2 \ell_2+3)^3}(\ell_2\pm\ell_1+1) (\ell_2\pm\ell_1+2)}\\
c_{02}^{\pm-}&=\pm\frac{i}{2}\sqrt{\frac{2 \ell_2+1}{2 \ell_2-1}(\ell_2\mp\ell_1) (\ell_2\mp\ell_1-1)}\\
c_{03}^{0+}&=-i (\ell_2 (\ell_2+2)-\lambda )\sqrt{\frac{1}{(2 \ell_2+1) (2 \ell_2+3)^3}(\ell_2-\ell_1+1) (\ell_2+\ell_1+1)}\\
c_{03}^{0-}&=i \sqrt{\frac{2 \ell_2+1}{2 \ell_2-1}(\ell_2-\ell_1) (\ell_2+\ell_1)}\\
c_{12}^{00}&=\ell_1\\
c_{13}^{\pm0}&=-\frac{1}{2} \sqrt{(\ell_2\mp\ell_1) (\ell_2\pm\ell_1+1)}\\
c_{23}^{\pm0}&=\mp\frac{i}{2}\sqrt{(\ell_2\mp\ell_1) (\ell_2\pm\ell_1+1)}
\end{aligned}
\end{equation*}

For $d=4$,
\begin{equation*}
\begin{aligned}
c_{01}^{\pm++}&=-\frac{1}{8} (\ell_3 (\ell_3+3)-\lambda )\cdot\\
&\hspace{10pt}\cdot\sqrt{\frac{1}{(\ell_3+1) (\ell_3+2)^3}\frac{(\ell_3+\ell_2+2) (\ell_3+\ell_2+3)}{(2 \ell_2+1) (2 \ell_2+3) }(\ell_2\pm\ell_1+1) (\ell_2\pm\ell_1+2) }\\
c_{01}^{\pm-+}&=-\frac{1}{8} (\ell_3 (\ell_3+3)-\lambda )\cdot\\
&\hspace{10pt}\cdot\sqrt{\frac{1}{(\ell_3+1) (\ell_3+2)^3}\frac{(\ell_3-\ell_2+2) (\ell_3-\ell_2+1)}{(2 \ell_2-1) (2 \ell_2+1) }(\ell_2\mp\ell_1) (\ell_2\mp\ell_1-1) }\\
c_{01}^{\pm+-}&=-\frac{1}{2} \sqrt{\frac{\ell_3+1}{\ell_3}\frac{ (\ell_3-\ell_2) (\ell_3-\ell_2-1)}{(2 \ell_2+1) (2 \ell_2+3) } (\ell_2\pm\ell_1+1) (\ell_2\pm\ell_1+2)}\\
c_{01}^{\pm--}&=-\frac{1}{2} \sqrt{\frac{\ell_3+1}{\ell_3}\frac{  (\ell_3+\ell_2) (\ell_3+\ell_2+1)}{(2 \ell_2-1) (2 \ell_2+1) }(\ell_2\mp\ell_1) (\ell_2\mp\ell_1-1)}\\
c_{02}^{\pm++}&=\mp\frac{i}{8} (\ell_3 (\ell_3+3)-\lambda )\cdot\\
&\hspace{10pt}\cdot\sqrt{\frac{1}{(\ell_3+1) (\ell_3+2)^3}\frac{ (\ell_3+\ell_2+2) (\ell_3+\ell_2+3)}{(2 \ell_2+1) (2 \ell_2+3) }(\ell_2\pm\ell_1+1) (\ell_2\pm\ell_1+2)}\\
c_{02}^{\pm-+}&=\mp\frac{i}{8} (\ell_3 (\ell_3+3)-\lambda )\cdot\\
&\hspace{10pt}\cdot\sqrt{\frac{1}{(\ell_3+1) (\ell_3+2)^3}\frac{(\ell_3-\ell_2+1) (\ell_3-\ell_2+2)}{(2 \ell_2-1) (2 \ell_2+1) }(\ell_2\mp\ell_1) (\ell_2\mp\ell_1-1) }\\
c_{02}^{\pm+-}&=\mp\frac{i}{2} \sqrt{\frac{\ell_3+1}{\ell_3}\frac{  (\ell_3-\ell_2) (\ell_3-\ell_2-1)}{(2 \ell_2+1) (2 \ell_2+3) }(\ell_2\pm\ell_1+1) (\ell_2\pm\ell_1+2)}\\
c_{02}^{\pm--}&=\mp\frac{i}{2} \sqrt{\frac{\ell_3+1}{\ell_3}\frac{  (\ell_3+\ell_2) (\ell_3+\ell_2+1)}{(2 \ell_2-1) (2 \ell_2+1) }(\ell_2\mp\ell_1) (\ell_2\mp\ell_1-1)}
\end{aligned}
\end{equation*}

\begin{equation*}
\begin{aligned}
c_{03}^{0++}&=+\frac{i}{4} (\ell_3 (\ell_3+3)-\lambda )\cdot\\
&\hspace{10pt}\cdot\sqrt{\frac{1}{ (\ell_3+1) (\ell_3+2)^3}\frac{ (\ell_3+\ell_2+2) (\ell_3+\ell_2+3)}{(2 \ell_2+1) (2 \ell_2+3)}(\ell_2-\ell_1+1) (\ell_2+\ell_1+1)}\\
c_{03}^{0-+}&=-\frac{i}{4} (\ell_3 (\ell_3+3)-\lambda )\cdot\\
&\hspace{10pt}\sqrt{\frac{1}{(\ell_3+1) (\ell_3+2)^3}\frac{ (\ell_3-\ell_2+1)(\ell_3-\ell_2+2) }{(2 \ell_2-1) (2 \ell_2+1) }(\ell_2-\ell_1) (\ell_2+\ell_1)}\\
c_{03}^{0+-}&=+i \sqrt{\frac{\ell_3+1}{\ell_3}\frac{  (\ell_3-\ell_2) (\ell_3-\ell_2-1)}{(2 \ell_2+1) (2 \ell_2+3) }(\ell_2-\ell_1+1) (\ell_2+\ell_1+1)}\\
c_{03}^{0--}&=-i \sqrt{\frac{\ell_3+1}{ \ell_3}\frac{  (\ell_2+\ell_3) (\ell_2+\ell_3+1)}{(1-2 \ell_2) (2 \ell_2+2-1)}(\ell_1-\ell_2) (\ell_1+\ell_2)}\\
c_{04}^{00+}&=-\frac{i}{4} (\ell_3 (\ell_3+3)-\lambda )\sqrt{\frac{}{(\ell_3+1) (\ell_3+2)^3}(\ell_3-\ell_2+1) (\ell_3+\ell_2+2)}\\
c_{04}^{00-}&=i \sqrt{\frac{\ell_3+1}{\ell_3}(\ell_3-\ell_2) (\ell_3+\ell_2+1)}\\
c_{14}^{\pm+0}&=\frac{1}{2} \sqrt{\frac{ (\ell_3-\ell_2) (\ell_3+\ell_2+2)}{(2 \ell_2+1) (2 \ell_2+3)}(\ell_2\pm\ell_1+1) (\ell_2\pm\ell_1+2)}\\
c_{14}^{\pm-0}&=\frac{1}{2} \sqrt{\frac{ (\ell_3-\ell_2+1) (\ell_3+\ell_2+1)}{(2 \ell_2-1) (2 \ell_2+1)}(\ell_2\mp\ell_1) (\ell_2\mp\ell_1-1)}\\
c_{24}^{\pm+0}&=\pm\frac{i}{2}  \sqrt{\frac{ (\ell_3-\ell_2) (\ell_3+\ell_2+2)}{(2 \ell_2+1) (2 \ell_2+3)}(\ell_2\pm\ell_1+1) (\ell_2\pm\ell_1+2)}\\
c_{24}^{\pm-0}&=\pm\frac{i}{2} \sqrt{\frac{ (\ell_3-\ell_2+1) (\ell_3+\ell_2+1)}{(2 \ell_2-1) (2 \ell_2+1)}(\ell_2\mp\ell_1) (\ell_2\mp\ell_1-1)}\\
c_{34}^{0+0}&=-i \sqrt{\frac{ (\ell_3-\ell_2) (\ell_3+\ell_2+2)}{(2 \ell_2+1) (2 \ell_2+3)}(\ell_2-\ell_1+1) (\ell_2+\ell_1+1)}\\
c_{34}^{0-0}&=+i \sqrt{\frac{ (\ell_3-\ell_2+1) (\ell_3+\ell_2+1)}{(2 \ell_2-1) (2 \ell_2+1)}(\ell_2-\ell_1) (\ell_2+\ell_1)}
\end{aligned}
\end{equation*}

For $d=5$,
\begin{equation*}
\begin{aligned}
c^{\pm+++}_{01}&=\frac{1}{4} (\ell_4 (\ell_4+4)-\lambda )\cdot\\
&\hspace{-5pt}\cdot\sqrt{\frac{1}{(2 \ell_4+3)(2 \ell_4+5)^3}\frac{(\ell_4+\ell_3+3) (\ell_4+\ell_3+4)}{(\ell_3+1) (\ell_3+2) }\frac{ (\ell_3+\ell_2+2) (\ell_3+\ell_2+3)}{(2 \ell_2+1) (2 \ell_2+3)  }(\ell_2\pm\ell_1+1) (\ell_2\pm\ell_1+2)}\\
c^{\pm+-+}_{01}&=\frac{1}{4} (\ell_4 (\ell_4+4)-\lambda )\cdot\\
&\hspace{-5pt}\cdot\sqrt{\frac{1}{(2 \ell_4+3) (2 \ell_4+5)^3}\frac{(\ell_4-\ell_3+1) (\ell_4-\ell_3+2)}{\ell_3 (\ell_3+1)}\frac{  (\ell_3-\ell_2) (\ell_3-\ell_2-1) }{(2 \ell_2+1) (2 \ell_2+3)  }(\ell_2\pm\ell_1+1)(\ell_2\pm\ell_1+2)}\\
c^{\pm-++}_{01}&=\frac{1}{4} (\ell_4 (\ell_4+4)-\lambda )\cdot\\
&\hspace{-5pt}\cdot\sqrt{\frac{1}{(2 \ell_4+3) (2 \ell_4+5)^3}\frac{(\ell_4+\ell_3+3) (\ell_4+\ell_3+4)}{ (\ell_3+1) (\ell_3+2)}\frac{ (\ell_3-\ell_2+2) (\ell_3-\ell_2+1) }{(2 \ell_2-1) (2 \ell_2+1) }(\ell_2\mp\ell_1) (\ell_2\mp\ell_1-1)}\\
c^{\pm--+}_{01}&=\frac{1}{4} (\ell_4 (\ell_4+4)-\lambda )\cdot\\
&\hspace{-5pt}\cdot\sqrt{\frac{1}{(2 \ell_4+3) (2 \ell_4+5)^3}\frac{ (\ell_4-\ell_3+2) (\ell_4-\ell_3+1)}{\ell_3 (\ell_3+1)}\frac{ (\ell_3+\ell_2) (\ell_3+\ell_2+1)}{(2 \ell_2-1) (2 \ell_2+1)  }(\ell_2\mp\ell_1) (\ell_2\mp\ell_1-1)}\\
c^{\pm+--}_{01}&=\frac{1}{4} \sqrt{\frac{2 \ell_4+3}{2 \ell_4+1}\frac{(\ell_4+\ell_3+1) (\ell_4+\ell_3+2)}{\ell_3 (\ell_3+1)}\frac{ (\ell_3-\ell_2) (\ell_3-\ell_2-1) }{(2 \ell_2+1) (2 \ell_2+3)  }(\ell_2\pm\ell_1+1) (\ell_2\pm\ell_1+2)}\\
c^{\pm++-}_{01}&=\frac{1}{4} \sqrt{\frac{2 \ell_4+3}{2 \ell_4+1}\frac{(\ell_4-\ell_3) (\ell_4-\ell_3-1)}{ (\ell_3+1) (\ell_3+2)}\frac{(\ell_3+\ell_2+2) (\ell_3+\ell_2+3) }{(2 \ell_2+1) (2 \ell_2+3) }(\ell_2\pm\ell_1+1) (\ell_2\pm\ell_1+2)}\\
c^{\pm-+-}_{01}&=\frac{1}{4} \sqrt{\frac{2 \ell_4+3}{2 \ell_4+1}\frac{(\ell_4-\ell_3) (\ell_4-\ell_3-1)}{ (\ell_3+1) (\ell_3+2)}\frac{(\ell_3-\ell_2+2) (\ell_3-\ell_2+1) }{(2 \ell_2-1) (2 \ell_2+1)}(\ell_2\pm\ell_1) (\ell_2\pm\ell_1-1)}\\
c^{\pm---}_{01}&=\frac{1}{4} \sqrt{\frac{2 \ell_4+3}{2 \ell_4+1}\frac{(\ell_4+\ell_3+1) (\ell_4+\ell_3+2)}{\ell_3 (\ell_3+1)}\frac{  (\ell_3+\ell_2) (\ell_3+\ell_2+1) }{(2 \ell_2-1) (2 \ell_2+1)  }(\ell_2\mp\ell_1) (\ell_2\mp\ell_1-1)}
\end{aligned}
\end{equation*}

\begin{equation*}
\begin{aligned}
c_{02}^{\pm+++}&=\pm\frac i4 (\ell_4 (\ell_4+4)-\lambda )\cdot\\
&\hspace{-5pt}\cdot\sqrt{\frac{1}{(2 \ell_4+3) (2 \ell_4+5)^3}\frac{ (\ell_4+\ell_3+3) (\ell_4+\ell_3+4)}{(\ell_3+1) (\ell_3+2)}\frac{ (\ell_3+\ell_2+2) (\ell_3+\ell_2+3)}{(2 \ell_2+1) (2 \ell_2+3)  }(\ell_2\pm\ell_1+1) (\ell_2\pm\ell_1+2)}\\
c_{02}^{\pm-++}&=\pm\frac i4(\ell_4 (\ell_4+4)-\lambda )\cdot\\
&\hspace{-5pt}\cdot\sqrt{\frac{1}{(2 \ell_4+3) (2 \ell_4+5)^3}\frac{(\ell_4+\ell_3+3) (\ell_4+\ell_3+4)}{ (\ell_3+1) (\ell_3+2)}\frac{ (\ell_3-\ell_2+2) (\ell_3-\ell_2+1) }{(2 \ell_2-1) (2 \ell_2+1)  }(\ell_2\mp\ell_1) (\ell_2\mp\ell_1-1)}\\
c_{02}^{\pm--+}&=\pm\frac i4(\ell_4 (\ell_4+4)-\lambda )\cdot\\
&\hspace{-5pt}\cdot\sqrt{\frac{1}{ (2 \ell_4+3)(2 \ell_4+5)^3}\frac{(\ell_4-\ell_3+1)(\ell_4-\ell_3+2)}{\ell_3 (\ell_3+1)}\frac{ (\ell_3+\ell_2) (\ell_3+\ell_2+1)  }{(2 \ell_2-1) (2 \ell_2+1)   }(\ell_2\mp\ell_1) (\ell_2\mp\ell_1-1)}\\
c_{02}^{\pm+-+}&=\pm\frac i4(\ell_4 (\ell_4+4)-\lambda )\cdot\\
&\hspace{-5pt}\cdot\sqrt{\frac{1}{(2 \ell_4+3)(2 \ell_4+5)^3 }\frac{(\ell_4-\ell_3+2) (\ell_4-\ell_3+1)}{\ell_3 (\ell_3+1)}\frac{ (\ell_3-\ell_2) (\ell_3-\ell_2-1) }{(2 \ell_2+1) (2 \ell_2+3)   }(\ell_2\pm\ell_1+1) (\ell_2\pm\ell_1+2)}\\
c_{02}^{\pm---}&=\pm \frac i4 \sqrt{\frac{2 \ell_4+3}{2 \ell_4+1}\frac{(\ell_4+\ell_3+1) (\ell_4+\ell_3+2)}{\ell_3 (\ell_3+1)}\frac{(\ell_3+\ell_2) (\ell_3+\ell_2+1) }{(2 \ell_2-1) (2 \ell_2+1) }(\ell_2\mp\ell_1) (\ell_2\mp\ell_1-1)}\\
c_{02}^{\pm+--}&=\pm \frac i4 \sqrt{\frac{2 \ell_4+3}{2 \ell_4+1}\frac{(\ell_4+\ell_3+1) (\ell_4+\ell_3+2)}{\ell_3 (\ell_3+1)}\frac{ (\ell_3-\ell_2) (\ell_3-\ell_2-1) }{(2 \ell_2+1) (2 \ell_2+3)  }(\ell_2\pm\ell_1+1) (\ell_2\pm\ell_1+2)}\\
c_{02}^{\pm++-}&=\pm \frac i4 \sqrt{\frac{2 \ell_4+3}{2 \ell_4+1}\frac{(\ell_4-\ell_3) (\ell_4-\ell_3-1)}{(\ell_3+1) (\ell_3+2)}\frac{(\ell_3+\ell_2+2) (\ell_3+\ell_2+3) }{(2 \ell_2+1) (2 \ell_2+3) } (\ell_2\pm\ell_1+1) (\ell_2\pm\ell_1+2)}\\
c_{02}^{\pm-+-}&=\pm \frac i4 \sqrt{\frac{2 \ell_4+3}{2 \ell_4+1}\frac{(\ell_4-\ell_3) (\ell_4-\ell_3-1)}{(\ell_3+1) (\ell_3+2)}\frac{ (\ell_3-\ell_2+2) (\ell_3-\ell_2+1) }{(2 \ell_2-1) (2 \ell_2+1) }(\ell_2\mp\ell_1) (\ell_2\mp\ell_1-1)}
\end{aligned}
\end{equation*}

\begin{equation*}
\begin{aligned}
c_{03}^{0+++}&=-\frac{i}{2}(\ell_4 (\ell_4+4)-\lambda )\cdot\\
&\hspace{-5pt}\cdot\sqrt{\frac{1}{ (2 \ell_4+3) (2 \ell_4+5)^3}\frac{ (\ell_4+\ell_3+3) (\ell_4+\ell_3+4)}{(\ell_3+1) (\ell_3+2)}\frac{ (\ell_3+\ell_2+2) (\ell_3+\ell_2+3)}{(2 \ell_2+1) (2 \ell_2+3) }(\ell_2-\ell_1+1) (\ell_2+\ell_1+1)}\\
c_{03}^{0-++}&=+\frac{i}{2} (\ell_4 (\ell_4+4)-\lambda )\cdot\\
&\hspace{-5pt}\cdot\sqrt{\frac{1}{(2 \ell_4+3) (2 \ell_4+5)^3}\frac{(\ell_4+\ell_3+3) (\ell_4+\ell_3+4)}{ (\ell_3+1) (\ell_3+2) }\frac{(\ell_3-\ell_2+2) (\ell_3-\ell_2+1) }{(2 \ell_2+1) (2 \ell_2-1) }(\ell_2-\ell_1) (\ell_2+\ell_1)}\\
c_{03}^{0+-+}&=-\frac{i}{2}(\ell_4 (\ell_4+4)-\lambda )\cdot\\
&\hspace{-5pt}\cdot\sqrt{\frac{1}{(2 \ell_4+3) (2 \ell_4+5)^3}\frac{ (\ell_4-\ell_3+2) (\ell_4-\ell_3+1)}{\ell_3 (\ell_3+1)}\frac{ (\ell_3-\ell_2) (\ell_3-\ell_2-1)}{(2 \ell_2+1) (2 \ell_2+3)  }(\ell_2-\ell_1+1) (\ell_2+\ell_1+1)}\\
c_{03}^{0--+}&=+\frac{i}{2} (\ell_4 (\ell_4+4)-\lambda )\cdot\\
&\hspace{-5pt}\cdot\sqrt{\frac{1}{ (2 \ell_4+3)(2 \ell_4+5)^3}\frac{ (\ell_4-\ell_3+2) (\ell_4-\ell_3+1)}{\ell_3 (\ell_3+1)}\frac{ (\ell_3+\ell_2) (\ell_3+\ell_2+1)}{(2 \ell_2-1) (2 \ell_2+1)}(\ell_2-\ell_1) (\ell_2+\ell_1)}\\
c_{03}^{0++-}&=-\frac{i}{2}\sqrt{\frac{2 \ell_4+3}{2 \ell_4+1}\frac{(\ell_4-\ell_3) (\ell_4-\ell_3-1)}{(\ell_3+1) (\ell_3+2)}\frac{ (\ell_3+\ell_2+2) (\ell_3+\ell_2+3)}{(2 \ell_2+1) (2 \ell_2+3)}(\ell_2-\ell_1+1) (\ell_2+\ell_1+1)}\\
c_{03}^{0-+-}&=+\frac{i}{2} \sqrt{\frac{2 \ell_4+3}{2 \ell_4+1}\frac{(\ell_4-\ell_3) (\ell_4-\ell_3-1)}{(\ell_3+1) (\ell_3+2)}\frac{(\ell_3-\ell_2+2) (\ell_3-\ell_2+1) }{(2 \ell_2+1) (2 \ell_2-1)}(\ell_2-\ell_1) (\ell_2+\ell_1)}\\
c_{03}^{0+--}&=-\frac{i}{2} \sqrt{\frac{2 \ell_4+3}{2 \ell_4+1}\frac{ (\ell_4+\ell_3+1) (\ell_4+\ell_3+2)}{\ell_3 (\ell_3+1)}\frac{(\ell_3-\ell_2) (\ell_3-\ell_2-1)}{(2 \ell_2+1) (2 \ell_2+3) }(\ell_2-\ell_1+1) (\ell_2+\ell_1+1)}\\
c_{03}^{0---}&=+\frac{i}{2} \sqrt{\frac{2 \ell_4+3}{2 \ell_4+1}\frac{(\ell_4+\ell_3+1) (\ell_4+\ell_3+2)}{\ell_3 (\ell_3+1)}\frac{(\ell_3+\ell_2) (\ell_3+\ell_2+1) }{(2 \ell_2-1) (2 \ell_2+1) }(\ell_2-\ell_1) (\ell_2+\ell_1)}
\end{aligned}
\end{equation*}

\begin{equation*}
\begin{aligned}
c_{04}^{00++}&=+\frac{i}{2} (\ell_4 (\ell_4+4)-\lambda )\cdot\\
&\hspace{10pt}\cdot\sqrt{\frac{1}{(2 \ell_4+3) (2 \ell_4+5)^3}\frac{ (\ell_4+\ell_3+3) (\ell_4+\ell_3+4)}{(\ell_3+1) (\ell_3+2)  }(\ell_3-\ell_2+1) (\ell_3+\ell_2+2)}\\
c_{04}^{00-+}&=-\frac{i}{2}(\ell_4 (\ell_4+4)-\lambda )\cdot\\
&\hspace{10pt}\cdot\sqrt{\frac{1}{ (2 \ell_4+3) (2 \ell_4+5)^3}\frac{ (\ell_4-\ell_3+2) (\ell_4-\ell_3+1)}{\ell_3 (\ell_3+1)}(\ell_3-\ell_2) (\ell_3+\ell_2+1)}\\
c_{04}^{00--}&=-\frac i2 \sqrt{\frac{2 \ell_4+3}{2 \ell_4+1}\frac{ (\ell_4+\ell_3+1) (\ell_4+\ell_3+2)}{\ell_3 (\ell_3+1)}(\ell_3-\ell_2) (\ell_3+\ell_2+1)}\\
c_{04}^{00+-}&=+\frac i2 \sqrt{\frac{2 \ell_4+3}{2 \ell_4+1}\frac{(\ell_4-\ell_3) (\ell_4-\ell_3-1)}{(\ell_3+1) (\ell_3+2)}(\ell_3-\ell_2+1) (\ell_3+\ell_2+2)}\\
c_{05}^{000+}&=-i (\ell_4 (\ell_4+4)-\lambda )\sqrt{\frac{1}{(2 \ell_4+3) (2 \ell_4+5)^3}(\ell_4-\ell_3+1) (\ell_4+\ell_3+3)}\\
c_{05}^{000-}&=+i \sqrt{\frac{2 \ell_4+3}{2 \ell_4+1}(\ell_4-\ell_3) (\ell_4+\ell_3+2)}\\
c_{15}^{\pm++0}&=-\frac14\sqrt{\frac{(\ell_4-\ell_3) (\ell_4+\ell_3+3)}{(\ell_3+1) (\ell_3+2)}\frac{ (\ell_3+\ell_2+2) (\ell_3+\ell_2+3) }{(2 \ell_2+1) (2 \ell_2+3) }(\ell_2\pm\ell_1+1) (\ell_2\pm\ell_1+2)}\\
c_{15}^{\pm--0}&=-\frac14\sqrt{\frac{(\ell_4-\ell_3+1) (\ell_4+\ell_3+2)}{\ell_3 (\ell_3+1)}\frac{ (\ell_3+\ell_2) (\ell_3+\ell_2+1) }{(2 \ell_2-1) (2 \ell_2+1)  }(\ell_2\mp\ell_1) (\ell_2\mp\ell_1-1)}\\
c_{15}^{\pm-+0}&=-\frac14\sqrt{\frac{(\ell_4-\ell_3) (\ell_4+\ell_3+3)}{ (\ell_3+1) (\ell_3+2)}\frac{(\ell_3-\ell_2+2) (\ell_3-\ell_2+1) }{(2 \ell_2-1) (2 \ell_2+1) }(\ell_2\mp\ell_1) (\ell_2\mp\ell_1-1) }\\
c_{15}^{\pm+-0}&=-\frac14\sqrt{\frac{(\ell_4-\ell_3+1) (\ell_4+\ell_3+2)}{\ell_3 (\ell_3+1)}\frac{ (\ell_3-\ell_2) (\ell_3-\ell_2-1) }{(2 \ell_2+1) (2 \ell_2+3)  }(\ell_2\pm\ell_1+1) (\ell_2\pm\ell_1+2)}
\end{aligned}
\end{equation*}

\begin{equation*}
\begin{aligned}
c_{25}^{\pm++0}&=\mp\frac i4 \sqrt{\frac{ (\ell_4-\ell_3) (\ell_4+\ell_3+3)}{(\ell_3+1) (\ell_3+2)}\frac{ (\ell_3+\ell_2+2) (\ell_3+\ell_2+3)}{(2 \ell_2+1) (2 \ell_2+3)  }(\ell_2\pm\ell_1+1) (\ell_2\pm\ell_1+2)}\\
c_{25}^{\pm--0}&=\mp\frac i4 \sqrt{\frac{(\ell_4-\ell_3+1) (\ell_4+\ell_3+2)}{\ell_3 (\ell_3+1)}\frac{ (\ell_3+\ell_2) (\ell_3+\ell_2+1) }{(2 \ell_2-1) (2 \ell_2+1)  }(\ell_2\mp\ell_1) (\ell_2\mp\ell_1-1)}\\
c_{25}^{\pm-+0}&=\mp\frac i4 \sqrt{\frac{ (\ell_4-\ell_3) (\ell_4+\ell_3+3)}{(\ell_3+1) (\ell_3+2)}\frac{ (\ell_3-\ell_2+2) (\ell_3-\ell_2+1)}{(2 \ell_2-1) (2 \ell_2+1)  }(\ell_2\mp\ell_1) (\ell_2\mp\ell_1-1)}\\
c_{25}^{\pm+-0}&=\mp\frac i4 \sqrt{\frac{(\ell_4-\ell_3+1) (\ell_4+\ell_3+2)}{\ell_3 (\ell_3+1)}\frac{ (\ell_3-\ell_2) (\ell_3-\ell_2-1) }{(2 \ell_2+1) (2 \ell_2+3)  }(\ell_2\pm\ell_1+1) (\ell_2\pm\ell_1+2)}\\
c_{35}^{0++0}&=+\frac i2 \sqrt{\frac{ (\ell_4-\ell_3) (\ell_4+\ell_3+3)}{(\ell_3+1) (\ell_3+2)}\frac{ (\ell_3+\ell_2+2) (\ell_3+\ell_2+3)}{(2 \ell_2+1) (2 \ell_2+3) }(\ell_2-\ell_1+1) (\ell_2+\ell_1+1)}\\
c_{35}^{0-+0}&=-\frac i2 \sqrt{\frac{(\ell_4-\ell_3) (\ell_4+\ell_3+3)}{(\ell_3+1) (\ell_3+2)}\frac{(\ell_3-\ell_2+2) (\ell_3-\ell_2+1) }{(2 \ell_2-1) (2 \ell_2+1) }(\ell_2-\ell_1) (\ell_2+\ell_1)}\\
c_{35}^{0+-0}&=+\frac i2 \sqrt{\frac{(\ell_4-\ell_3+1) (\ell_4+\ell_3+2)}{\ell_3 (\ell_3+1)}\frac{ (\ell_3-\ell_2) (\ell_3-\ell_2-1) }{(2 \ell_2+1) (2 \ell_2+3)  }(\ell_2-\ell_1+1) (\ell_2+\ell_1+1)}\\
c_{35}^{0--0}&=-\frac i2 \sqrt{\frac{ (\ell_4-\ell_3+1) (\ell_4+\ell_3+2)}{\ell_3 (\ell_3+1)}\frac{ (\ell_3+\ell_2) (\ell_3+\ell_2+1)}{(2 \ell_2-1) (2 \ell_2+1)  }(\ell_2-\ell_1) (\ell_2+\ell_1)}\\
c_{45}^{00+0}&=-\frac i2 \sqrt{\frac{ (\ell_4-\ell_3) (\ell_4+\ell_3+3)}{ (\ell_3+1) (\ell_3+2)}(\ell_3-\ell_2+1) (\ell_3+\ell_2+2)}\\
c_{45}^{00-0}&=+\frac i2 \sqrt{\frac{ (\ell_4-\ell_3+1) (\ell_4+\ell_3+2)}{ \ell_3 (\ell_3+1)}(\ell_3-\ell_2) (\ell_3+\ell_2+1)}
\end{aligned}
\end{equation*}

%

\restoregeometry

\bibliographystyle{ieeetr}
\bibliography{references}

\end{document}